\documentclass[11pt,a4paper]{article}
\pdfoutput=1
\usepackage{graphicx,epstopdf}
\usepackage{soul}
\usepackage{natbib}
\usepackage{amsmath,amsfonts,amssymb,amsbsy,amsthm,mathtools, mathrsfs}
\usepackage{bbm,bm}
\usepackage[usenames,dvipsnames]{color} 
\usepackage{xcolor}
\usepackage{subfig}
\usepackage{booktabs}
\usepackage{url}
\usepackage{eurosym}
\usepackage[a4paper]{geometry}
\usepackage{a4wide}
\usepackage[colorlinks]{hyperref}
\usepackage{authblk}
\usepackage[linesnumbered,ruled,vlined]{algorithm2e}
\allowdisplaybreaks
\numberwithin{equation}{section}
\usepackage{enumitem,float}
\usepackage{fancyvrb}
\usepackage[utf8]{inputenc}
\usepackage{dsfont}
\usepackage{orcidlink,thumbpdf,lmodern}
\usepackage{arydshln}
\usepackage[normalem]{ulem}

\DeclareMathOperator*{\argmin}{arg\,min}

\newcommand{\diff}{\mathrm{d}}

\DontPrintSemicolon
\newcommand{\To}{\mbox{\upshape\bfseries to}}

\newcommand{\bzero}{\bm{0}}

\def\H0{{\mathcal{H}_0}}

\def\Lik{{\mathcal{L}}}

\def\MDA{\mathcal{D}}

\renewcommand{\leq}{\leqslant}
\renewcommand{\geq}{\geqslant}

\hypersetup{colorlinks,%
citecolor=blue,%
filecolor=green,%
linkcolor=red,%
urlcolor=violet,%
}

\usepackage{listings}
\definecolor{backcolour}{rgb}{0.95,0.95,0.92}
\definecolor{codegreen}{rgb}{0,0.6,0}

\lstdefinestyle{myStyle}{
    language=R,
    backgroundcolor=\color{white},   
    commentstyle=\color{ForestGreen},
    stringstyle=\color{Purple},
    keywordstyle=\color{RoyalBlue},
    basicstyle=\ttfamily\footnotesize,
    frame=single, 
    rulecolor=\color{black},
    breakatwhitespace=false,         
    breaklines=true,                 
    keepspaces=true,                 
    numbers=left,       
    numbersep=5pt,                  
    showspaces=false,                
    showstringspaces=false,
    showtabs=false,                  
    tabsize=2,
}

\newlist{inparaenum}{enumerate}{2}
\setlist[inparaenum,1]{label=(\roman*)}
\setlist[inparaenum,2]{label=(\roman{inparaenumi}\emph{\alph*})}

\newcommand{\bfa}{{\boldsymbol{a}}}
\newcommand{\bfb}{{\boldsymbol{b}}}

\newcommand{\bfr}{{\boldsymbol{r}}}

\newcommand{\bfu}{{\boldsymbol{u}}}
\newcommand{\bfv}{{\boldsymbol{v}}}
\newcommand{\bfw}{{\boldsymbol{w}}}
\newcommand{\bfx}{{\boldsymbol{x}}}
\newcommand{\bfy}{{\boldsymbol{y}}}
\newcommand{\bft}{{\boldsymbol{t}}}
\newcommand{\bfz}{{\boldsymbol{z}}}
\newcommand{\bfeta}{{\boldsymbol{\eta}}}
\newcommand{\bfalpha}{{\boldsymbol{\alpha}}}
\newcommand{\bfmu}{{\boldsymbol{\mu}}}
\newcommand{\bfsigma}{{\boldsymbol{\sigma}}}
\newcommand{\bfbeta}{{\boldsymbol{\beta}}}
\newcommand{\bflambda}{{\boldsymbol{\lambda}}}
\newcommand{\bfvartheta}{{\boldsymbol{\vartheta}}}
\newcommand{\bftheta}{{\boldsymbol{\theta}}}
\newcommand{\bfgamma}{{\boldsymbol{\gamma}}}
\newcommand{\bfvarphi}{{\boldsymbol{\varphi}}}

\newcommand{\bfM}{{\boldsymbol{M}}}

\newcommand{\bfR}{{\boldsymbol{R}}}

\newcommand{\bfX}{{\boldsymbol{X}}}

\newcommand{\bfY}{{\boldsymbol{Y}}}

\def\real{{\mathbb R}}

\def\bzero{{\bf 0}}

\def\indic{\mathds{1}}
\def\MDA{\mathcal{D}}
\newcommand{\resimp}{{\mathcal{R}}}
\def\simp{\mathcal{S}}
\newcommand{\evc}{{C_{\scalebox{0.65}{EV}}}}
\def\prob{{\mathbb{P}}}
\def\expect{{\mathbb{E}}}
\newcommand{\pickset}{\mathscr{A}}

\def\Lik{{\mathcal{L}}}

\makeatletter
\def\adl@drawiv#1#2#3{%
        \hskip.5\tabcolsep
        \xleaders#3{#2.5\@tempdimb #1{1}#2.5\@tempdimb}%
                #2\z@ plus1fil minus1fil\relax
        \hskip.5\tabcolsep}
\newcommand{\cdashlinelr}[1]{%
  \noalign{\vskip\aboverulesep
           \global\let\@dashdrawstore\adl@draw
           \global\let\adl@draw\adl@drawiv}
  \cdashline{#1}
  \noalign{\global\let\adl@draw\@dashdrawstore
           \vskip\belowrulesep}}
\makeatother

\let\proglang=\textsf
\newcommand{\pkg}[1]{{\fontseries{m}\fontseries{b}\selectfont #1}}
\let\code=\texttt
\DefineVerbatimEnvironment{CodeInput}{Verbatim}{fontshape=sl}
\DefineVerbatimEnvironment{CodeOutput}{Verbatim}{}
\newenvironment{CodeChunk}{}{}

 \title{Modeling extremal dependence in multivariate and spatial problems: a practical perspective}  
   
\author[1~\orcidlink{0000-0002-7944-3925}]{Boris B\'{e}ranger}
\affil[1]{\footnotesize \emph{School of Mathematics and Statistics, UNSW Data Science Hub (uDASH), UNSW Sydney}
\authorcr \emph{E-mail:} \url{B.Beranger@unsw.edu.au}}
\author[2~\orcidlink{0000-0002-0417-7570}]{Simone A. Padoan}
\affil[2]{\footnotesize \emph{Department of Decision Sciences, Bocconi University} \authorcr
\emph{E-mail:} \url{Simone.Padoan@unibocconi.it}} 

\begin{document}

\maketitle
   
\begin{abstract}
From environmental sciences to finance, there is a growing demand for methods that can assess the risks of extreme events beyond those observed in available data. Extrapolating extreme events beyond the range of the data is not obvious. 
Risk assessments are often further complicated by the need to account for multiple variables simultaneously. Extreme value theory provides important tools for the analysis of multivariate or spatial extreme events, but these are not easily accessible to professionals without appropriate expertise. This article provides a minimal background on multivariate and spatial extremes and gives simple yet thorough instructions on how to analyse them using the \proglang{R} package \pkg{ExtremalDep}. After briefly introducing the statistical methodologies, we focus on road testing the package's toolbox through several real-world applications.
\end{abstract}

\noindent\emph{Keywords:} Angular measure, Air pollution, Environmental data analysis, Exchange rates, Extremal coefficient, Extreme sets, Heat Waves, Max-stable processes, Multivariate generalized extreme-value distribution, Pickands dependence function, Quantile regions



%
\section[Introduction]{Introduction}
 In many applied fields—including environmental sciences, finance, and actuarial sciences—it is essential to assess the risk of extreme events beyond those observed in historical data, with the goal to anticipate the magnitude of hypothetical future catastrophes (e.g., a global financial crisis or large monetary losses due
to natural hazards). The aim is to disclose the uncertainty surrounding such extreme events to decision makers in order to carefully plan mitigation strategies. A key challenge arises when extrapolating beyond the range of $N$ available observations, by computing for example quantiles whose probability ($p$) is smaller than $1/N$, meaning that the largest data observation is not expected to exceed the target event on average. This makes direct empirical estimation impossible. Extreme Value Theory (EVT) addresses this problem by providing a rigorous probabilistic framework and statistical methodology for modeling and extrapolating extreme events. For accessible introductions, see, for instance, \citet{beirlant2004}, \citet{de2006}, and \citet{falk2010}.

In the univariate case, the theory and practice is mature with a significant number of packages available on the Comprehensive \proglang{R} Archive Network (CRAN), forming an extensive set of tools. Examples include \pkg{ismev} \citep{ismev}, \pkg{evd} \cite{Stephenson2002}, \pkg{revdbayes} \citep{revdbayes}, \pkg{extRemes} \citep{extRemes} and \pkg{ExtremeRisks} \citep{extremerisks} to name a few. 
In practice, extreme events rarely involve a single variable. Multiple factors are typically observed simultaneously: for example, rainfall, temperature, and wind measured across different locations within a region can jointly drive economic losses from adverse weather. Similarly, high pollution levels often result from the combined effect of several pollutants exceeding safety thresholds.

When it comes to multivariate and spatial problems, statistical methodologies for analyzing multivariate and spatial extremes, and their corresponding software implementations, remain relatively underdeveloped. The dependence structure governing multivariate extreme value distributions, or equivalently, the finite-dimensional distributions of stochastic processes for extremes is termed extremal dependence.
It is an infinite-dimensional, nonparametric object subject to some restrictive conditions, which leads to both a complex formulation of the overall multivariate distribution and significant inferential challenges. The literature on multivariate and spatial extremes covers several topics including multivariate extremes based on componentwise maxima \citep[e.g.,][]{Beranger2015} and multiple peaks over threshold \citep[e.g.,][]{rootzen2018multivariate}, spatial extremes based on pointwise maxima and peaks over threshold \citep[e.g.,][]{davison2012statistical, davison2018}, graphical models \citep{graphical2020} and time series \citep[e.g.,][]{Extremogram}, and remains nowadays a vibrant research field with a number of ongoing new results. These have produced useful \proglang{R} packages such as \pkg{mev} \citep{mev}, \pkg{SpatialExtremes} \citep{SpatialExtremes210}, \pkg{mvPot} \citep{mvpot}, \pkg{RandomFields}, \pkg{graphicalExtremes} \citep{graphicalextremes}, \pkg{extremogram} \citep{extremogrampkg} as well as \pkg{ExtremalDep} \citep{extremaldep}. A review of statistical software for modeling extremes is provided by \citet{Belzile2023}, which introduces some of the capabilities of \pkg{ExtremalDep}; but we offer here a more in-depth guide to the package, complemented by practical illustrations.

 The focus of this work is on modeling the extremal dependence structure arising from the componentwise maxima approach, which enables the assessment of risk associated with multivariate extreme events in a broad sense. Contrary to what might be expected, this framework is not limited to modeling multivariate maxima, but is also well suited for analyzing multivariate threshold exceedances. This is achieved by exploiting approximations of the probabilities of observing at least one exceedance \eqref{eq:stable_tail_approx}, joint exceedances \eqref{eq:tail copula_tail_approx}, or an observation falling within an extreme region \eqref{eq:quantile_region_prob}, in which at least one component is extreme. Statistical methods for risk analysis based on these probability approximations are presented in Sections \ref{subsec:parametric_model}, \ref{sssec:sim_ev}, and \ref{sssec:exQ}.
If the primary interest lies instead in proper multivariate peaks-over-threshold models and methods specifically tailored to their analysis, we refer the reader to \citet{rootzen2018multivariate, davison2018,  kiriliouk2019peaks}. Finally, it is worth noting that several alternative approaches for modeling multivariate extremes also exist, including conditional extremes models \citep{heffernan2004}, random scale and location mixture models \citep{huser2017bridging}, and (X-)vine copulae \citep{kiriliouk2025x}. Several of these alternative methods are implemented in the \pkg{texmex} package \citep{texmex}.

 Most existing packages for the analysis of multivariate extremes rely on parametric representations of extremal dependence (e.g., \pkg{mev}, \pkg{mvPot}, and \pkg{graphicalExtremes}), which simplify the underlying mathematics and inference but impose strong structural assumptions. While \pkg{ExtremalDep} also accommodates parametric approaches, one of its main contributions lies in extending the available toolkit by providing nonparametric and semiparametric methods that remain closer to the inherently nonparametric nature of extremal dependence.
A further distinguishing feature of \pkg{ExtremalDep}, compared with existing software, is its emphasis on demonstrating how extremal dependence can be directly exploited for risk assessment related to multivariate extreme events and their occurrence probabilities. In particular, the routines of the package allow one to compute joint return levels and conditional return periods (Section \ref{subsec:parametric_model}), conditional probabilities given the occurrence of extreme events (Section \ref{sssec:beed_np}), tail probabilities relative to multiple threshold exceedances (Section \ref{sssec:sim_ev}), and extreme quantile regions (Section \ref{sssec:exQ}). Finally, the package allows for frequentist and Bayesian inference for multivariate extremes and frequentist and Bayesian inference for spatial extremes.
This manuscript demonstrates the practical use of the package through several important applications. 
Because the ordering of points in $d$-dimensional Euclidean space is arbitrary, we introduce and formalize key concepts such as extreme sets, multivariate extreme quantiles, quantile regions, joint tail and conditional probabilities, and multivariate return levels.
In practice, these tools enable a wide range of risk analyses. Examples include risk assessment of future high-pollution episodes in urban areas, identifying regions of a country exposed to heavy precipitation events, or evaluating the risk of future heatwave episodes. 
The article is divided into two parts: Section \ref{sec:multi_extremes} discusses multivariate extremes while Section \ref{sec_spa_extremes} focuses on spatial extremes.
In the multivariate part, Section \ref{sec:back_ground} gives a brief introduction to the theoretical background while Section \ref{subsec:parametric_model} details how to perform parametric (frequentist and Bayesian) inference for the extremal dependence and for joint and conditional tail probabilities.
Section \ref{subsec:nonparametric_model} explains how to perform nonparametric Bayesian inference for the extremal dependence of a $d$-dimensional componentwise maxima vector, and once the extremal dependence has been estimated, how to infer joint and conditional tail probabilities in the bivariate case. The focus then successively shifts to the random generation of bivariate extreme values with a flexible semi-parametric dependence structure, the estimation of small probabilities of belonging to certain extreme sets and the estimation extreme quantile regions corresponding to small joint tail probabilities. The paper ends in Section \ref{sec_conclusion} with a summary and a discussion on future developments.

 %
\section[Multivariate extremes]{Multivariate extremes}\label{sec:multi_extremes}
%
%
\subsection[Background]{Background}\label{sec:back_ground}
Let $\bfX_1,\bfX_2,\ldots$ be independent and identically distributed (i.i.d.) random vectors, where $\bfX_i=(X_{i,1},\ldots,X_{i,d})\in\real^d$ for some integer $d>1$, whose joint distribution $F$ is in the domain of attraction of a \textit{multivariate extreme value} distribution $G$ \citep[chap. 6]{de2006}, denoted by $F\in\MDA(G)$. This implies that  for $n=1,2,\ldots$, there exists norming sequences $\bfa_n>\bzero_d=(0,\ldots,0)$ where the inequality is applied element-wise, and $\bfb_n\in\real^d$ such that
\begin{equation}\label{eq:MDOA}
\lim_{n\to\infty} F^n(\bfa_n \bfx + \bfb_n)=G_\bfgamma(\bfx\mid D), \quad \bfx\in\real^d,
\end{equation}
where $\bfgamma=(\gamma_1,\ldots,\gamma_d)^\top \in \real^d$.
The limiting distribution in \eqref{eq:MDOA} takes the form
\begin{equation}\label{eq:evd_copula}
G_{\bfgamma}(\bfx\mid D) = \evc \left( G_{\gamma_1}(x_1), \ldots, G_{\gamma_d}(x_d) \mid D \right), \qquad \bfx \in \real^d,
\end{equation}
where $\evc$ is an \textit{extreme-value copula} \citep[chap. 8]{beirlant2004}, representing the so-called \textit{extremal dependence}, which binds the $d$ marginal distributions that are, for all $j\in\{1,\ldots,d\}$, univariate GEV distributions of the form
\begin{equation}\label{eq:uni_gev}
%
%
G_{\gamma_j}(x_j)=\exp\left\{-\left(1+\gamma_j x_j\right)^{-\frac{1}{\gamma_j}}\right\},\quad 1+\gamma_j x_j>0,
\end{equation}
where $\gamma_j\in\real$ is the {\it extreme value index}. 
This coefficient describes the tail heaviness of the distribution, resulting in a heavy-, light- or short-tailed distribution, depending on whether $\gamma_j>0$, $\gamma_j=0$ or $\gamma_j<0$ \citep[e.g.,][chap. 1]{de2006}. In \eqref{eq:evd_copula} the extreme-value copula is defined as
\begin{equation}\label{eq:evcopula}
\evc(\bfu \mid D) =
 \exp \left\{ - L\left(  - \ln u_1, \ldots, - \ln u_d\right) \right\}, 
\quad \bfu \in (0, 1]^d,
\end{equation}
where $L: [0,\infty)^d \mapsto [0,\infty)$ is a homogeneous function of order $1$ named {\it stable-tail} dependence function \citep[chap. 8]{beirlant2004}. This latter function fully characterizes the extremal dependence structure; however, since it is defined on $[0,\infty)^d$ and takes values in $[0,\infty)$, it is difficult to visualize directly. For this reason, it is common practice to work instead with its restriction to the set $\resimp := \{\bft \in [0,1]^{d-1} : \lVert \bft \rVert_1 \leq 1\}$, known as the Pickands dependence function \citep[e.g.,][Ch.~4]{falk2010}, which is more amenable to graphical representation. 
While no exhaustive characterization of the Pickands dependence function is available, it must at least satisfy the following convexity and boundary conditions
\begin{enumerate}
\item[(C1)] $A(a\bft_1+(1-a)\bft_2)\leq aA(\bft_1)+(1-a)A(\bft_2), \,a\in[0,1], \,\forall\,\bft_1,\bft_2\in\resimp$,
\item[(C2)] $1/d\leq \max\left(t_1,\ldots,t_{d-1},1-t_1-\cdots-t_{d-1} \right) \leq A(\bft) \leq 1, \,\forall\,\bft\in\resimp$.
\end{enumerate}
In the bivariate case, these are necessary and sufficient to define a valid Pickands dependence function \citep[for details, see][chap.~8]{beirlant2004}. However, there is an alternative way to represent extreme dependence. Indeed, a formal definition of the Pickands dependence function is
%
%
\begin{equation}\label{eq:picklands}
A(\bft)=L(1-t_1-\dots-t_{d-1}, \bft)=d\int_{\simp_d} \max\{(1-t_1-\cdots-t_{d-1})w_1, \ldots,  t_{d-1}w_d\} \diff H( \bfw),
\end{equation}
where $H$ is a probability measure on the $d$-dimensional unit simplex  ${\simp_d:=\{\bfw\geq\bzero: \Vert \bfw \Vert_1=1\}}$, named \textit{angular measure} \citep[e.g.,][Ch. 4]{falk2010}. Thus, the level of dependence can be conveyed through $H$; however, the same issue arises again: which specific measure should one consider? A valid angular measure is any probability measure satisfying the mean constraint 
\begin{enumerate}
\item[(C3)] $\int_{\simp_d} w_j \diff H(\bfw)=1/d,\; \forall\;j\in\{1,\ldots,d\}$.
\end{enumerate}
Since $A$ and $H$ are related through the one-to-one map in formula \eqref{eq:picklands}, the dependence parameter $D$ of the copula $\evc(\cdot\mid D)$ can be equivalently meant to be either the angular measure or the Pickands dependence function. Both are possible means to describe the extremal dependence.

%
\begin{figure}[t!]
\centering
$
\begin{array}{ccc}
\includegraphics[width=0.25\textwidth, page=2]{examples.pdf} 
\includegraphics[width=0.25\textwidth, page=1]{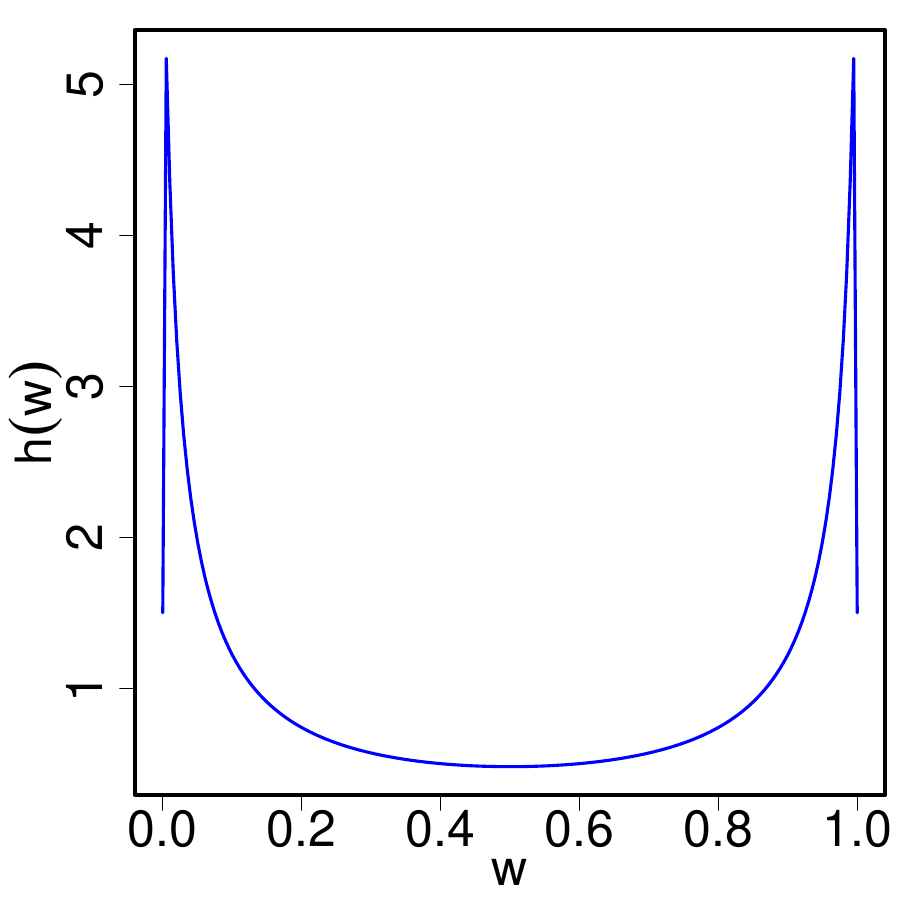} 
\includegraphics[width=0.25\textwidth, page=4]{examples.pdf}
\includegraphics[width=0.25\textwidth, page=3]{examples.pdf} 
\end{array}
$
\caption{\label{fig:examples} From left to right, the first two panels display angular densities corresponding to strong and weak dependence between two maxima, respectively. In the strong dependence case, the probability mass concentrates around $1/2$, whereas in the weak dependence case it concentrates near the vertices of $(0,1)$. The third and fourth panels show the corresponding Pickands dependence functions, which are close to the boundary cases 
$\max(w,1-w)$ and $1$, respectively.} 
\end{figure}
The result in \eqref{eq:MDOA} can be interpreted as follows.
Given a sample $(\bfX_1, \ldots, \bfX_{N})$ with joint distribution $F\in\MDA(G)$, for each of the $d$ variables we compute $k$ maxima are computed over blocks of size $n$, such that $N=n\cdot k$, where the $i$th maximum is $M_{i,j}^{(n)}=\max(X_{(i-1)n+1,j},\ldots,X_{in,j})$, $i\in\{1,\ldots,k\}$, $j=1,\ldots,d$. We bind the maxima in the vector of componentwise maxima $\bfM_{i}^{(n)}=(M_{i,1}^{(n)}, \ldots,M_{i,d}^{(n)})^\top$.
Now, setting $\bfmu=\bfb_n$, $\bfsigma=\bfa_n$, $\bfy=\bfa_n \bfx + \bfb_n$, then thanks to \eqref{eq:MDOA}, the joint distribution of vector of componentwise maxima can be approximated, for large enough $n$, as
$
F^n(\bfy)\approx G_{\bftheta}(\bfy\mid D),
$
where $\bftheta=(\bfmu,\bfsigma,\bfgamma)^\top$,  By construction, the distribution of the univariate maximum $M_{i,j}^{(n)}$ can be approximated by
\begin{equation}\label{eq:uniGEV}
F^n(y_j)\approx G_{\theta_j}(y_j)=
\exp\left\{-\left(1+\gamma_j \frac{y_j-\mu_j}{\sigma_j}\right)^{-\frac{1}{\gamma_j}}_+\right\}, 
\quad j\in\{1,\ldots,d\},
\end{equation}
where the dependence among the componentwise maxima $(M_{i,1}^{(n)}, \ldots, M_{i,d}^{(n)})$ can be approximated through the extreme-value copula $\evc(\cdot\mid D)$. When the dependence parameter $D$ is represented by the angular measure $H$ the interpretation is as follows: the more mass $H$ assigns to the center of the simplex $(1/d, \ldots, 1/d)$, the stronger the dependence among the maxima; conversely, the dependence weakens as $H$ concentrates its mass near the vertices of the simplex. This behavior is illustrated by the solid lines in the first and second panels of Figure~\ref{fig:examples}. When $D$ is expressed in terms of the Pickands dependence function, dependence strength is reflected by its proximity to the bounds in condition (C2). Specifically, the closer the Pickands dependence function is to the lower bound, the stronger the dependence among the maxima, whereas proximity to the upper bound indicates weaker dependence. These contrasting regimes are illustrated by the solid lines in the third and fourth panels of Figure~\ref{fig:examples}.

A useful summary of the dependence among the maxima is given by the so-called {\it extremal coefficient}  \citep[e.g.,][Ch. 8]{beirlant2004}, defined as
\begin{equation}\label{eq:extremal_coefficient}
\varrho = d A(1/d,\ldots,1/d),
\end{equation}
which represents the (fractional) number of independent block maxima. The extremal coefficient satisfies $1\leq \varrho \leq d$, with the lower and upper bounds describing the case of complete dependence and independence.

Even more importantly, beyond these results, the domain of attraction provides also a useful basis to derive small joint tail probabilities corresponding to multivariate extremes relative to the original distribution $F$ such as multiple threshold exceedances and derive extreme quantile regions corresponding to some small joint tail probabilities. These results are summarized as follows.
The limit in \eqref{eq:MDOA} holds if and only if $n\{1-F(\bfa_n \bfx + \bfb_n)\}\to -\log G(\bfx)$ as $n\to\infty$, for all $\bfx\in\real^d$, with the convention that $-\log(0)=\infty$. Together with the weak convergence of the copula of the vector of componentwise maxima, namely $\{C(\bfu^{1/n})\}^n\to \evc(\bfu)$ as $n\to\infty$, for all $\bfu \in (0, 1]^d$, this result implies that for all $\bfx\in[0,\infty)^d$, as $n\to \infty$,
\begin{equation}
\label{eq:stable_tail}
n\,\prob\left(\bigcup_{j=1}^d F_j(X_j)> 1-\frac{x_j}{n}\right)\to 
L(\bfx)=d\int_{\simp_d} \max_{j=1,\ldots,d}{(x_j}w_j) \diff H( \bfw).
\end{equation}
Recall that the joint distribution function $F$ can be represented in terms of a copula function, $C$, defined on $[0,1]^d$ and its marginal distributions $F_i, i=1, \ldots, d$ (assumed to be continuous) through $F(\bfx)=C\{F_1(x_1),\ldots,F_d(x_d)\}$ \citep[see, e.g.,][for details]{joe2014dependence}.
The limiting result in formula \eqref{eq:stable_tail} implies the following important approximation: let $p_j=x_j/n$, for $j=1,\ldots,d$, and $Q_j(p_j)$ be the ($1-p$) quantile of $F_j$; 
then, by homogeneity of the stable tail dependence function, the probability that at least one among the $d$ variables $X_j$ exceeds a high quantile of its own distribution, can be approximated for large $n$ by
\begin{equation}\label{eq:stable_tail_approx}
\prob\{X_1>Q_1(p_1)\, \text{or} \cdots \text{or} \, X_d>Q_d(p_d)\}\approx L(p_1,\ldots,p_d).
\end{equation}
This result is particularly relevant in applications because, once the stable tail dependence function has been estimated from the data, for small probabilities $p_1,\ldots,p_d$, it directly provides an estimate of the probability that at least one of the $d$ variables of interest exceeds a high threshold.
By the inclusion and exclusion principle, the weak convergence result of the extreme value copula also allows one to obtain
\begin{equation*}\label{eq:tail copula_tail}
n\,\prob\left(\bigcap_{j=1}^d  F_j(X_j) > 1-\frac{x_j}{n}\right)\to R(\bfx)=d\int_{\simp_d} \min_{j=1,\ldots,d}{(x_j}w_j) \diff H( \bfw), \quad n\to \infty, 
\end{equation*}
where $R$ is known as the \textit{tail copula} function, which is also a homogeneous function of order $1$, and  therefore the probability that all the $d$ variables $(X_1,\ldots, X_d)$ exceed simultaneously high quantiles of their own distributions can be approximated, for a sufficiently large $n$, by
\begin{equation}\label{eq:tail copula_tail_approx}
\prob\{X_1>Q_1(p_1)\, \text{and} \cdots   \text{and}\, X_d>Q_d(p_d)\}\approx R(p_1,\ldots,p_d).
\end{equation}
Also this result is very important for practice, as once the tail copula function has been estimated, corresponding to small probabilities $p_1,\ldots,p_d$, it implicitly provides an estimate of the probability that all the $d$ variables of interest exceeds high thresholds.

Finally, an important question is, how to determine an extreme region where two variables, given a small joint probability $p$ of falling in it? We summarize the approach discussed by \citet{cai2011}, \cite{einmahl2013} and \citet{he2017} as follows with a  focus on the bivariate case explored by \cite{einmahl2013}. Let $f$ be the density on $\real_+^2$ of a bivariate distribution $F$ satisfying $F\in\MDA(G)$, with $\gamma_1,\gamma_2>0$. A possible definition of quantile regions is given by the level sets of $f$, namely for $\alpha>0$
\begin{equation}\label{eq:quantile_region}
\mathcal{Q}=\{\bfx\in\real^2_+:f(\bfx)\leq \alpha\}.
\end{equation}
Then, for a small probability $p>0$, we can derive the region $\mathcal{Q}$ such that $\prob(\mathcal{Q})=p$ and $f$ everywhere on $\mathcal{Q}$ is smaller than $f$  everywhere on $Q^{\complement}$. Therefore, the latter is the smallest region satisfying $\prob(\mathcal{Q}^{\complement})=1-p$. Accordingly, an {\it extreme quantile region} is defined by a level set $\mathcal{Q}_N$ (with $\alpha=\alpha_N$ in \ref{eq:quantile_region}), which is such that $\prob(\mathcal{Q}_N)=p$, where $p=p_N\to 0$ and the expected number of points falling in it is $Np\to c>0$ as $N\to\infty$. Note that when $c<1$, the region is so extreme that almost no observations are expected to fall in it. From a practical perspective, given a sample of large size $N$ and a tiny probability $p$ of falling within an extreme quantile region, how can such a region be determined?

The domain of attraction condition implies the existence of a measure $\xi$, named {\it exponent measure}, satisfying $\xi(\partial B)=0$, for all Borel set $B\subset[0,\infty]^2$ that are
bounded away from the origin, and such that the probability of falling in extreme region $N B$ can be approximated as
\begin{equation}\label{eq:quantile_region_prob}
\prob \left( \left\{ Q_1(1/N)x_1^{\gamma_1}, Q_2(1/N)x_2^{\gamma_2}:\bfx\in B\right\} \right)\approx\xi(NB), \quad\text{as}\quad N\to\infty.
\end{equation}
Assume that for some $\mathcal{M}>0$, $f$ is bounded away from zero on $(0,\mathcal{M}]^2$ and, outside this region, $f$ is decreasing in each coordinate and that the exponent measure admits a density $g$ satisfying 
$$
\xi(\{\bfv:v_1>x_1\, \text{ or } \,v_2>x_2\})=\int\int_{\{v_1>x_1 \text{ or } v_2>x_2\}} g(\bfv) \diff \bfv,
$$
and
$$
\lim_{N\to\infty}NQ_1(1/N)Q_2(1/N) f\left\{Q_1(1/N)x^{\gamma_1},Q_2(1/N)x^{\gamma_2}\right\}
=\frac{x_1^{1-\gamma_1}x_2^{1-\gamma_2}}{\gamma_1\gamma_2} g(\bfx)=:q(\bfx).
$$
Note that $g$ is related to the density $h$ of the angular measure $H$ by $h(w)=2^{-1}g(w,1-w)$, where $w=x_1/r$ and $r=x_1+x_2$. The main idea behind the derivation of an extreme quantile region is, first to define the so-called \textit{basic set} $S$ defined as 
$$S=\{\bfx:q(\bfx)\leq 1\}=\{\bfx:r\geq q_\star^{-1}(w), w\in[0,1]\},$$
 by means of the \textit{angular basic density}
$$
q_\star(w)=\left\{2\gamma_1^{-1}\gamma_2^{-1} w^{1-\gamma_1}(1-w)^{1-\gamma_2}h(w)\right\}^{-\frac{1}{1+\gamma_1+\gamma_2}}.
$$
According to the exponent measure, its size is equal to 
$$
\xi(S)=2\int_{[0,1]}q_{\star}(w)h(w)\diff w.
$$
Then, the basic set $S$ is suitably inflated into an extreme set $\widetilde{\mathcal{Q}}_N$ defined as
\begin{equation}\label{eq:app_extreme_region}
\widetilde{\mathcal{Q}}_N=\left\{\left(\mu_1+\sigma_1\frac{\left(\frac{k\,\xi(S)x_1}{Np}\right)^{\gamma_1}-1}{\gamma_1},\mu_2+\sigma_2\frac{\left(\frac{k\,\xi(S)x_2}{Np}\right)^{\gamma_2}-1}{\gamma_2}\right):\bfx\in S\right\},
\end{equation}
where $k=k_N$ with $k\to\infty$ as $N\to\infty$ and $k=o(N)$, which  approximate the target extreme quantile region $\mathcal{Q}_N$.
In particular, $\mu_j=Q_j(k/N)$ and $\sigma_j=a_j(N/k)$ with $j=1,2$, where $a_j(\cdot)$ is a suitable positive scale function \citep[][Ch. 1--2]{de2006}. Then, the set $\widetilde{\mathcal{Q}}_N$ is a good approximation of $\mathcal{Q}_N$ in the sense that $\prob(\widetilde{\mathcal{Q}}_N)\approx p$ and $\prob(\mathcal{Q}_N\Delta \widetilde{\mathcal{Q}}_N)\to0$ as  $N\to\infty$ \citep[see][]{einmahl2013}, where $A\Delta B=A \setminus B \cup B \setminus A$.

\subsection[Statistical challenging]{Statistical challenges}

From a statistical perspective, the multivariate GEV model defines a complex semiparametric family of distributions of the form $G_{\bftheta}(\bfy\mid D)=\evc\{G_{\theta_1}(y_1),\ldots,G_{\theta_d}(y_d)\mid D\}$, $\bfy\in\real^d$, where $\bftheta\in\Theta\subset (\real,\real_+,\real)^d$ is a finite-dimensional vector of marginal parameters and $D$ is an infinite-dimensional dependence parameter  of the copula $\evc$, named extremal dependence, living in a suitable space, depending whether $D$ is the angular measure or the Pickands dependence function. The estimation of such infinite-dimensional parameter, accounting for the corresponding constraints (see conditions (C1)--(C3) of Section \ref{sec:back_ground}), is far from simple. 
To reduce complexity, many classes of parametric models were proposed to model $D$, see  \cite{Beranger2015} for a review. In order to provide valid extremal dependence structures, the proposed models satisfy the required constraints. 
Section \ref{subsec:parametric_model} describes how to fit some of the most popular parametric models for $D$ using \pkg{ExtremalDep}.

Over time, more sophisticated semi- and non-parametric approaches for modeling and estimating the extremal dependence have been proposed, under both parametrisations. Some examples include: projection estimators \citep[e.g.,][]{fils2008}, polynomials  \citep[e.g.,][]{Marcon2017a} and splines \citep[e.g.,][]{cormier2014}, just to name a few. Section \ref{subsec:nonparametric_model} details how to use  \pkg{ExtremalDep} for non-parametric estimation of the extremal dependence, simulation of bivariate extreme values with a flexible semi-parametric dependence structure, estimation of small probabilities to belong to certain extreme sets and estimation extreme quantile regions corresponding to small joint tail probabilities.
\subsection[Parametric modeling of the extremal dependence]{Parametric modeling of the extremal dependence}\label{subsec:parametric_model}

\noindent\textbf{Theory}. The Poisson Point Process (PPP) is a useful characterization for statistical models used to represent the extremal dependence, when the latter is parametrized according to the angular measure and provided that it allows for a density $h(w\mid\bfvarphi)$ on $\simp_d$, where $\bfvarphi$ is a suitable vector of parameters \citep[see, e.g.,][]{coles1991, engelke2015}. In particular, let $\bfx_1,\ldots,\bfx_N$ be i.i.d. observations from $F\in\MDA(G)$ and $\bfy_1,\ldots,\bfy_N$ be their equivalent with unit Fr\'{e}chet margins. The pseudopolar transformation $T$ is defined as
$T(\bfy_i)=(r_{(i)},\bfw_{(i)})$ where $r^{(i)}=y_{i,1}+\cdots+y_{i,d}$ is the radial part and $\bfw_{(i)}=\bfy_i/ r_{(i)}$ is the angular part of $\bfy_i$, $i=1,\ldots,N$. 
The domain of attraction condition implies that, if $r_0$ is a large threshold and $W_{r_0}=\{(r,\bfw):r>r_0\}$ is a set of points with a radial
component larger than $r_0$, the number of points falling in $W_{r_0}$ is described approximatively by a Poisson random variable with rate $1/\psi(W_{r_0})$, where $\psi=\xi\{T^{-1}(\cdot)\}$. Thus, conditional on observing $k$ points $\{(r_{(i)},\bfw_{(i)}),i=1,\ldots,k\}$ with radial part greater than $r_0$, such points are approximately independent with the density
 function $r^{-2}\diff r \times H(\diff \bfw)/\psi(W_{r_0})$. The likelihood function of the exceedances can then be approximated by
\begin{align*}
L(\bfvarphi\mid(r_{(i)},\bfw_{(i)}),i=1,\ldots, k)&\approx\frac{e^{-\psi(W_{r_0})}\{\psi(W_{r_0})\}^{k}}{k!}
\prod_{i=1}^k\frac{r_{(i)}^{-2}h(\bfw_{(i)}\mid\bfvarphi)}{\psi(W_{r_0})}\\
&\propto h(\bfw_{(i)}\mid\bfvarphi).
\end{align*}
Estimates of $\bfvarphi$ can be obtained by maximizing the log-likelihood $\ell(\bfvarphi)\propto\sum_{i=1}^k \log h(\bfw_{(i)}\mid\bfvarphi)$. Alternatively, one can appeal to a  Bayesian approach \citep{sabourin2013}, based on the approximate posterior density of the angular density parameters given by
$$
\pi(\bfvarphi\mid\bfw_{(1)},\ldots,\bfw_{(k)})=\frac{\prod_{i=1}^k h(\bfw_{(i)}\mid\bfvarphi)\phi(\bfvarphi)}{\int_{\Psi} \prod_{i=1}^k h(\bfw_{(i)}\mid\bfvarphi)\phi(\bfvarphi)\diff \bfvarphi},
$$
where $\phi(\bfvarphi)$ is a suitable prior distribution on $\bfvarphi$ and $\Psi$ is a suitable parameter space.\\

\noindent\textbf{Software implementation}. A non-exhaustive list of popular parametric models for the extremal dependence includes the
asymmetric logistic (\texttt{AL}), pairwise beta (\texttt{PB}), tilted Dirichlet (\texttt{TD}), H\"{u}sler-Reiss (\texttt{HR}), extremal-$t$ (\texttt{ET}) and extremal-skew-$t$ (\texttt{EST}) models \citep[see e.g.,][]{Beranger2015, Beranger2017}. The routine \code{fExtDep} with argument \code{method = "PPP"} allows to estimate $\bfvarphi$ by likelihood maximisation. The model class for $h$ is specified with the argument \code{model} with the list of available options reported in Table \ref{tab:prop_prior}. Some classes of models are characterised by an angular density $h(\bfw\mid\bfvarphi)$ that places a continuous mass only in the interior of $\simp_d$ (\code{PB}, \code{TD}, \code{HR}), while others also allow to place densities and atoms on subspaces of $\simp_d$. \citet{Beranger2017} describe an extended version of a Maximum Likelihood (ML) method that allows for the estimation of these more complex models up to the three-dimensional case (\code{AL}, \code{ET}, \code{EST}) by exploiting the following approach. 

\begin{itemize}
\item In the bivariate case, an observation $\bfw=(w,1-w)$ falls at an endpoint of $\simp_d$ if $w < c$ or $w > 1-c$ for some small $c \in (0,1/2)$ and on the interior of $\simp_d$ otherwise. 
\item A trivariate observation $\bfw=(w_1,w_2,1-w_1-w_2)$ falls: at the $j$th corner of $\simp_d$ if $\bfw \in \mathcal{C}_j = \left\{w_j > 1-c \right\} $, on the edge between the $j$th and $l$th components if $\bfw \in \mathcal{E}_{j,l} = \left\{w_j, w_l < 1-c, w_j + w_l > 1-c, w_j > 1- 2 w_l, w_l > 1 -2 w_j \right\}$, and in the interior of $\simp_d$ otherwise (i.e., if $w_j > c,$ for all $j$) \citep[see][for details]{Beranger2017}.
\end{itemize}

 This approach is implemented when specifying a positive value for the argument \code{c} of \code{fExtDep}.\\

\noindent\textbf{Application}. We illustrate the use of \code{fExtDep} on the analysis of high levels of air pollution recorded in Leeds, UK, over the winter period (November 1st to February 28/29th) between 1994 and 1998 \citep[see][]{heffernan2004}. The \code{pollution} dataset contains the daily maximum of the five air pollutants: particulate matter (PM10),
nitrogen oxide (NO), nitrogen dioxide (NO2), ozone (03), and sulfur dioxide (SO2),  on the unit Fr\'echet scale.  For a description of the transformation used see \cite{Beranger2015}. The datasets \code{PNS}, \code{NSN} and \code{PNN} contains the angular components corresponding to the 100 largest radial components of the triplets of pollutants (PM10, NO, SO2), (NO2, SO2, NO)  and (PM10, NO, NO2), respectively. 

We run the following commands
\begin{CodeChunk}
\begin{CodeInput}
R> data(pollution, package = "ExtremalDep")
R> f.et05 <- fExtDep(x = PNS, method = "PPP", model = "ET", 
+    par.start = c(0.5, 0.5, 0.5, 3), trace = 2, c = 0.05)
\end{CodeInput}
%\begin{CodeOutput}
%initial  value -118.269640 
%iter  10 value -146.907539
%iter  10 value -146.907539
%final  value -146.907539 
%converged
%\end{CodeOutput}
\begin{CodeInput}
R> f.et <- fExtDep(x = PNS, method = "PPP", model = "ET", 
+    par.start = c(0.5, 0.5, 0.5, 3), trace = 2 )
\end{CodeInput}
%\begin{CodeOutput}
%initial  value -161.078869 
%iter  10 value -225.654851
%iter  20 value -234.524086
%iter  30 value -235.285461
%iter  40 value -235.360084
%final  value -235.382293 
%converged
%\end{CodeOutput}
\end{CodeChunk}
where
the first call to the \code{fExtDep} routine fits the extremal-$t$ angular density (\code{model = "ET"}), to the pollution data (\code{x = PSN}), allowing for point masses at the corners by specifying  the argument \code{c}, while the second call considers a density defined only in the interior of $\simp_d$.
The argument \code{par.start} gives starting values for the parameters and \code{trace = 2} allows to monitor optimization progress. The optimization method is set by the argument \code{optim.method} (\code{"BFGS"} by default) on which basis the routine \code{optim} from the \code{stats} library is used.
When \code{method = "PPP"} or \code{"composite"}, \code{fExtDep} returns a list object of class \code{ExtDep\_Freq} from which the estimates of the model parameters, their standard errors, the log-likelihood maximum and the Takeuchi Information Criterion (TIC) can be extracted respectively with the \code{est}, \code{StdErr}, \code{logLik} and \code{tic} functions. The fitted model and estimation method can also be extracted using the \code{model} and \code{method} S3 generics. \code{fExtDep} is first ran for the extremal-$t$ model using different values of $c$ ($0, 0.05$ and $0.1$) to take into account point masses at the corners of  $\simp_d$. The TICs indicate that only assuming mass in the interior of $\simp_d$ provides the best results (lowest TIC). A performance comparison between the \code{PB}, \code{AL}, \code{TD}, \code{HR}, \code{ET} and \code{EST} models defined on the interior only, is performed with TIC scores reported in
Table~\ref{tab:TICmod} and as a result the extremal skew-$t$ model fits the data best (lowest TIC).
\begin{table}[h!]
\centering
\begin{tabular}{ccccccc}
\toprule
\code{model} & \code{PB} & \code{AL} & \code{TD} & \code{HR} & \code{ET}  & \code{EST}  \\
\midrule
\textrm{TIC}  & $-$188.75 & $-$406.85 & $-$393.75 & $-$460.25  & $-$461.74  & $-$493.31 \\
\bottomrule
\end{tabular}
\caption{TIC of models fitted to \code{PNS} data obtained using \code{fExtDep} with \code{c=0}. See Table~\ref{tab:prop_prior} for a description of the acronyms.}
\label{tab:TICmod}
\end{table}

\begin{figure}[t!]
\centering
$
\begin{array}{ccc}
\textrm{pairwise beta} & \textrm{asymmetric logistic} & \textrm{tilted Dirichlet} \\
\includegraphics[width=0.25\textwidth]{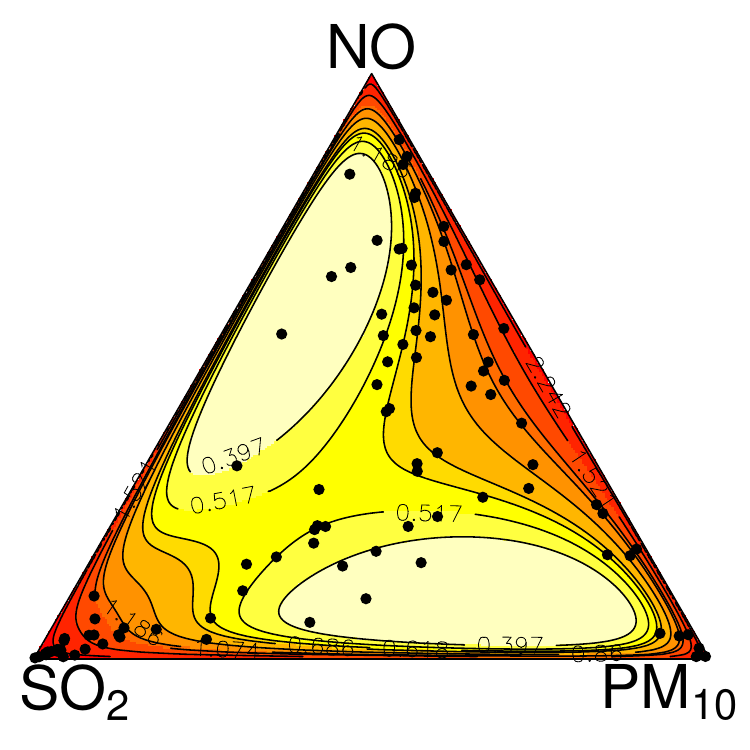} &
\includegraphics[width=0.25\textwidth]{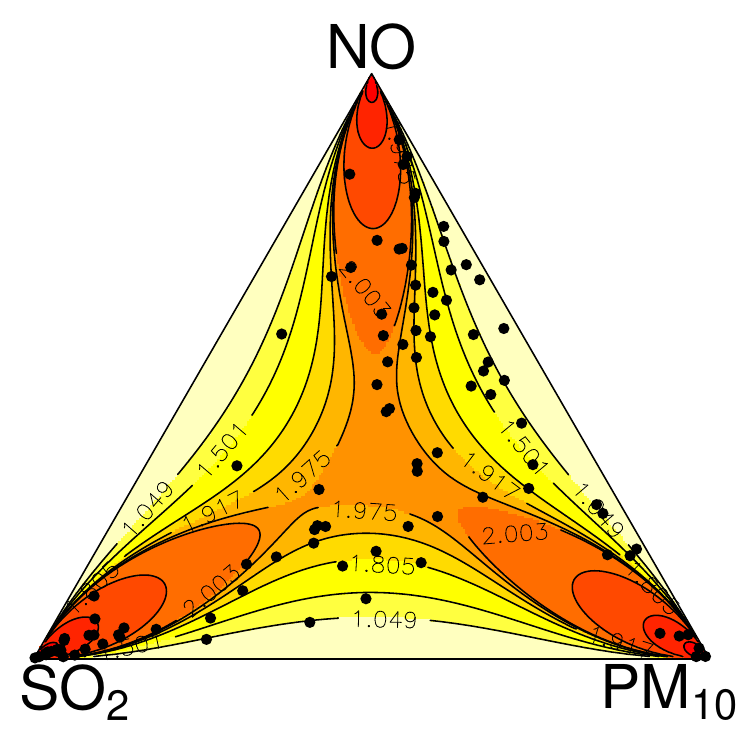} &
\includegraphics[width=0.25\textwidth]{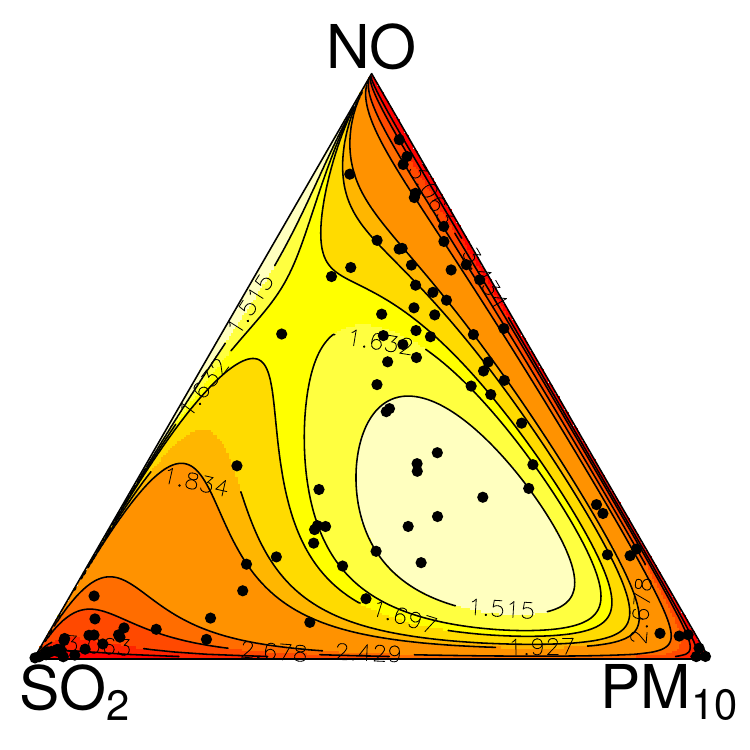} \\
 \textrm{H\"{u}sler-Reiss} & \textrm{extremal-$t$} & \textrm{extremal skew-$t$} \\
\includegraphics[width=0.25\textwidth]{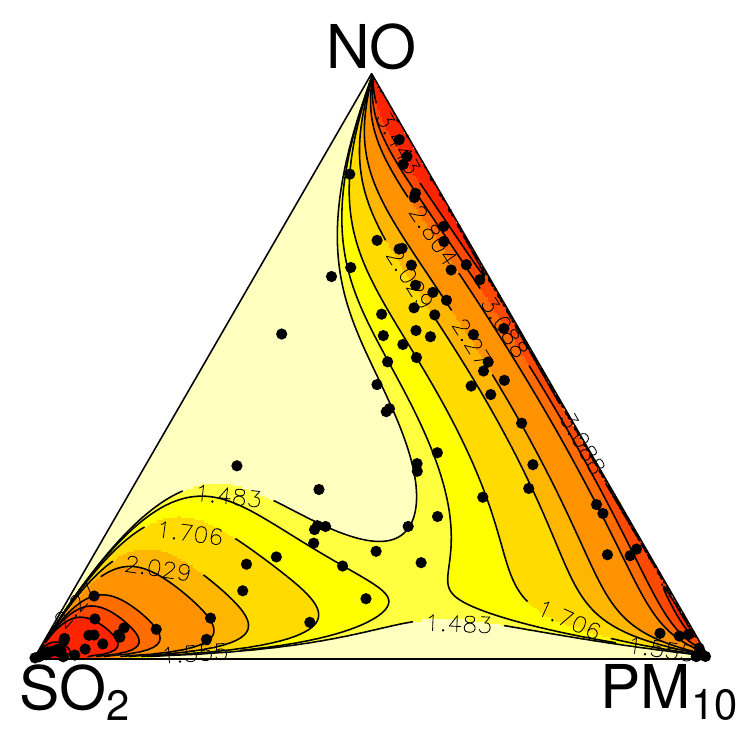} &
\includegraphics[width=0.25\textwidth]{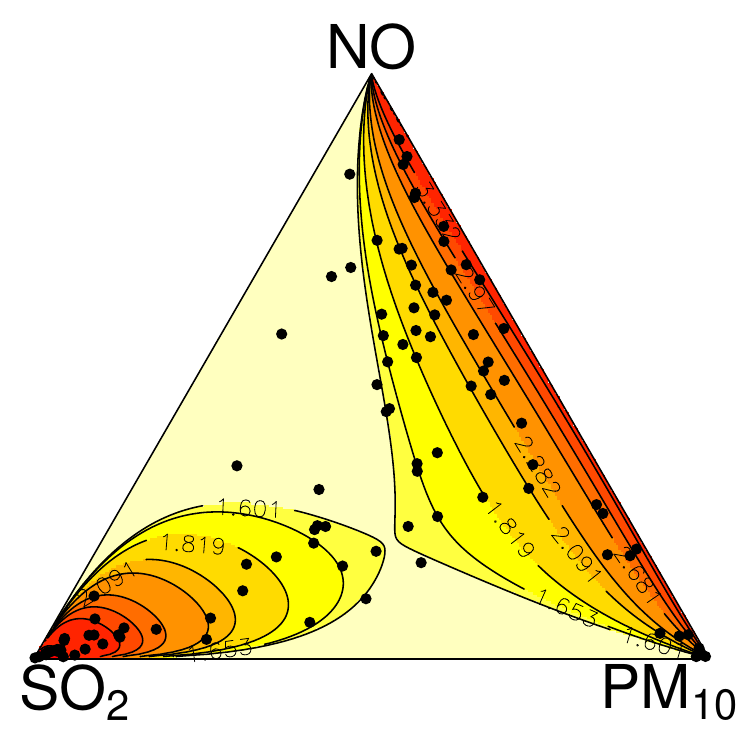} &
\includegraphics[width=0.25\textwidth]{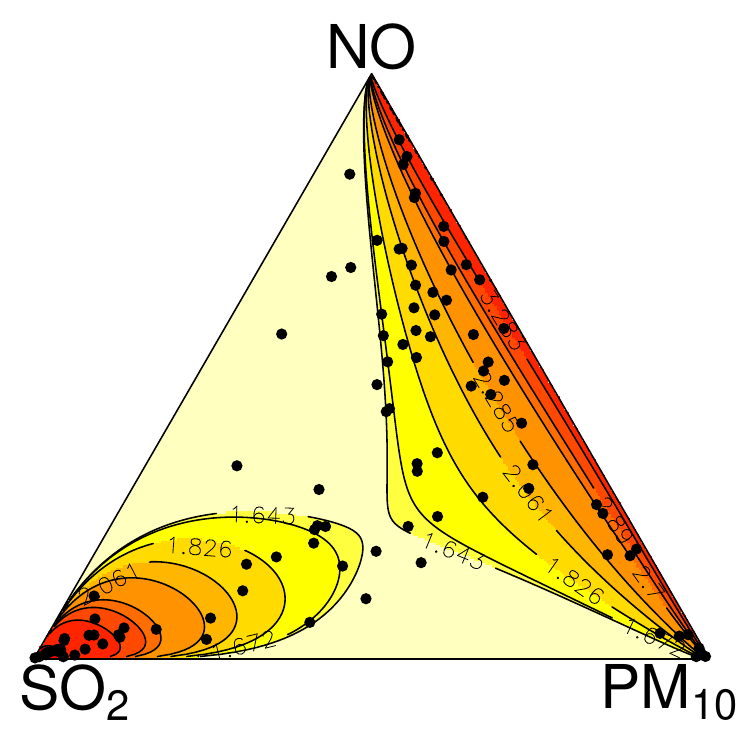} 
\end{array}
$
\caption{\label{fig:Angular_fits} Estimated angular densities of the models from Table \ref{tab:prop_prior}.}
\end{figure}

Bayesian inference for the parameter $\bfvarphi$ is implemented through a model averaging approach \citep{sabourin2013} and available in the routine \code{fExtDep} using \code{method = "BayesianPPP"}.  Table \ref{tab:prop_prior} reports the implemented prior densities for each angular density model. Let  $\phi(\cdot, a,b)$ denote a univariate Gaussian density with mean $a$ and variance $b$ and $t(\cdot)$ a transformation applied to the parameters to map them to the real line. For each component of $\bfvarphi$, the proposal density is Gaussian of the form $\phi\{t(\varphi_j^{\star}\}; t(\varphi_j^{(i)}),\tt{MCpar})$, where $\varphi_j^{\star}$ and $\varphi_j^{(i)}$ are the proposed value and the value at the $i$th iteration of the algorithm of the $j$ element of $\bfvarphi$, respectively, and \code{MCpar} is the variance term that needs to be specified by the user. 
\begin{CodeChunk}
\begin{CodeInput}
R> Hpar.hr <- list(mean.lambda = 0, sd.lambda = 3)
R> PNS.hr <- fExtDep(x = PNS, method = "BayesianPPP", model = "HR", 
+    Nsim = 3e3, Nbin = 1e3, Hpar = Hpar.hr, MCpar = 0.05, seed = 14342)
\end{CodeInput}
%\begin{CodeOutput}
%|=====================================================================| 100%
%\end{CodeOutput}
\end{CodeChunk}
The above display illustrates how to perform Bayesian inference for the H\"{us}sler--Reiss model (\code{model = "HR"}) with  3,000 iterations (\code{Nsim}), a burn-in period of 1,000 iterations (\code{Nbin}) while the hyper-parameters are set through \code{Hpar} and we set the random seed \code{seed} for reproducibility. Given that we have a Gaussian prior distribution for the model's parameter on the log scale, setting the hyper-parameters \code{mean.lambda=0} and \code{sd.lambda=3} provides relatively non-informative prior information about the parameter. 
When \code{method = "BayesianPPP"}, \code{fExtDep} returns an object of class \code{ExtDep\_Bayes} from which the posterior mean, posterior standard deviation and Bayesian Information Criterion can respectively be extracted using the \code{est}, \code{StdErr} and \code{bic} functions. Note that here we have $2,000$ posterior draws corresponding to \code{Nsim - Nbin}. Again, the \code{model} and \code{method} function can be used to extract the model name and estimation procedure. 
%
%
\begin{table}[b!]
\centering
\begin{tabular}{ccccl}
\toprule
\textrm{Model} &  \code{model} &  $\bfvarphi$ & $ t(\cdot)$ & \textrm{Prior} \\
\hline
pairwise beta & \code{PB} & $\left(\alpha, \beta_{i,j} \right)$ & $\log$ & $\phi$($\log(\alpha)$; {\tt mean.alpha}, {\tt sd.alpha}) \\
& & & $\log$ & $\phi$($\log(\beta_{i,j})$; {\tt mean.beta}, {\tt sd.beta}) \\
\cdashlinelr{1-5}
asymmetric logistic &  \code{AL} & $\left( \alpha_{\mathcal{S}}, \beta_{j, \mathcal{S}} \right)$ & $\log$ & $\phi\left( \log(\alpha_{\mathcal{S}} ); {\tt mean.alpha}, {\tt sd.alpha} \right)$ \\
& & & \textrm{logit} & $\phi\left(\textrm{logit}(\beta_{j,\mathcal{S}} ); {\tt mean.beta}, {\tt sd.beta} \right)$ \\
\cdashlinelr{1-5}
tilted Dirichlet &  \code{TD} &  $\alpha_{j} $ & $\log$ & $\phi(\log(\alpha_{j}); {\tt mean.alpha}, {\tt sd.alpha})$ \\ 
\cdashlinelr{1-5}
H\"{u}sler--Reiss &  \code{HR} &  $\lambda_{i,j}$ & $\log$ & $\phi(\log(\lambda_{i,j}); {\tt mean.lambda}, {\tt sd.lambda})$ \\ 
\cdashlinelr{1-5}
extremal-$t$ &   \code{ET} &  $ \left( \rho_{i,j}, \nu \right)$ & \textrm{atanh} & $\phi(\textrm{atanh}(\rho_{i,j}); {\tt mean.rho}, {\tt sd.rho})$ \\
 & & & $\log$ & $\phi(\log(\nu)$; {\tt mean.nu}, {\tt sd.nu}) \\ 
\cdashlinelr{1-5}
extremal-skew-$t$ &   \code{EST} &  $\left(\rho_{i,j}, \alpha_i, \nu \right)$ & \textrm{atanh} & $\phi(\textrm{atanh}(\rho_{i,j}); {\tt mean.rho}, {\tt sd.rho})$\\
&  & & \textrm{identity} & $\phi(\alpha_i; {\tt mean.alpha}, {\tt sd.alpha})$\\   
&  & & $\log$ & $\phi(\log(\nu); {\tt mean.nu}, {\tt sd.nu}) $\\ 
\bottomrule
\end{tabular}
\caption{List of available angular density model families and prior densities that are considered with them when using \code{fExtDep(method = "BayesianPPP", ...)}.}
\label{tab:prop_prior}
\end{table} %
The empirical mean (standard deviation) obtained from the approximate posterior distribution are $\hat{\lambda}_{1,2} = 0.65 (0.04), \hat{\lambda}_{1,3} = 0.90(0.04)$ and $\hat{\lambda}_{2,3} = 0.98(0.04)$, while the Bayesian Information Criterion (BIC) is $-449.65$. Visualization of the fitted angular density and data is obtained by calling the \code{plot} routine on the object of class \code{ExtDep\_Bayes} with \code{type = "angular"} as default.
\begin{CodeChunk}
\begin{CodeInput}
R> labs <- c(expression(PM[10]), expression(NO), expression(SO[2]))
R> plot(PNS.hr, labels = labs, cex.lab = 2.4)
\end{CodeInput}
\end{CodeChunk}
%
Figure~\ref{fig:Angular_fits} showcases the fitted dependence structures for the H\"{u}sler-Reiss model as well as the other models listed in Table \ref{tab:prop_prior}, together with the data (solid black dots). Note that the package ensures that the covariance matrix involved with the H\"{u}sler-Reiss model is non-negative definite. From visual inspection, the H\"{u}sler-Reiss, extremal-$t$ and extremal skew-$t$ model appear to follow the data structure well. Given that the former is the simplest model, we focus on the H\"{u}sler-Reiss model  for the remainder of this section.
%

\begin{figure}[t!]
\centering
$
\begin{array}{ccc}
\includegraphics[width=0.3\textwidth]{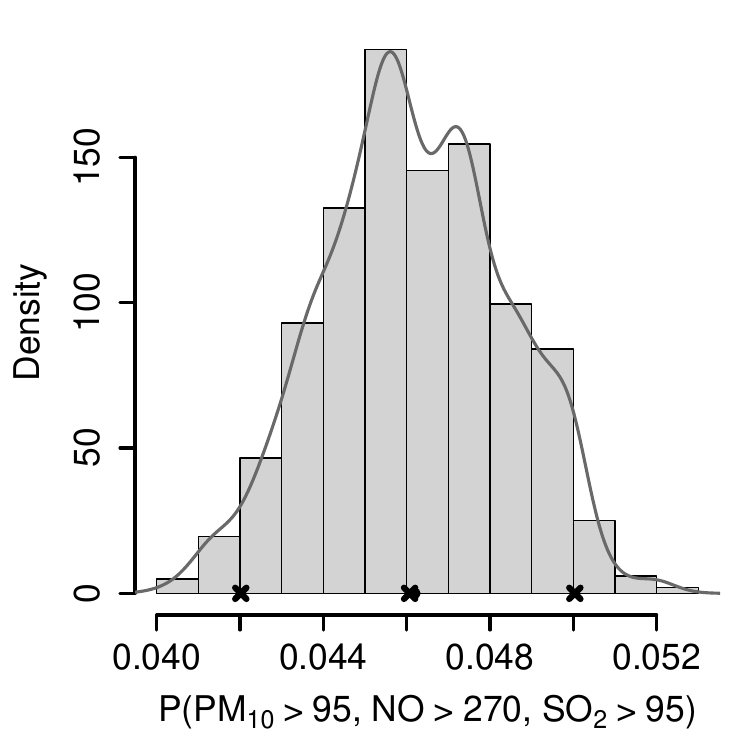} &
\includegraphics[width=0.3\textwidth]{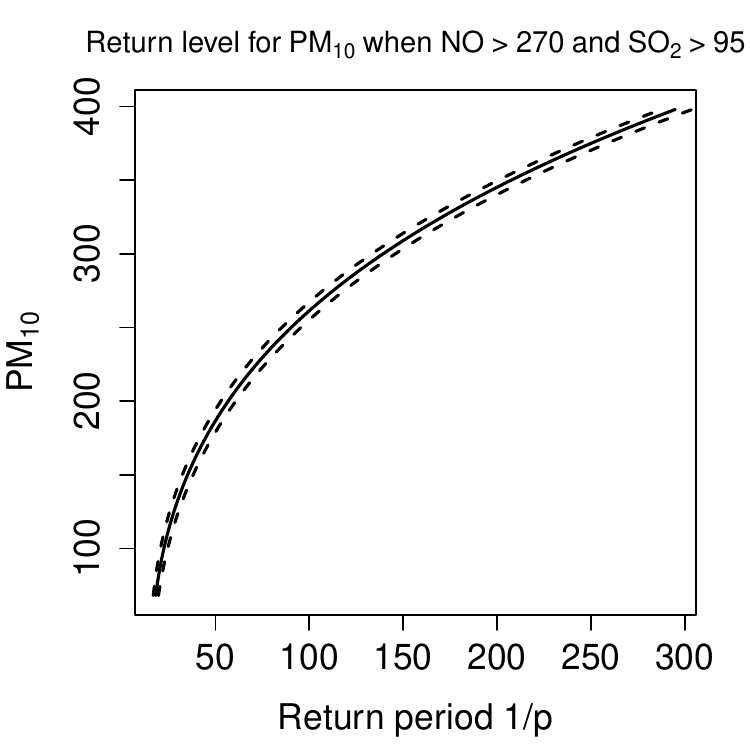} &
\includegraphics[width=0.3\textwidth]{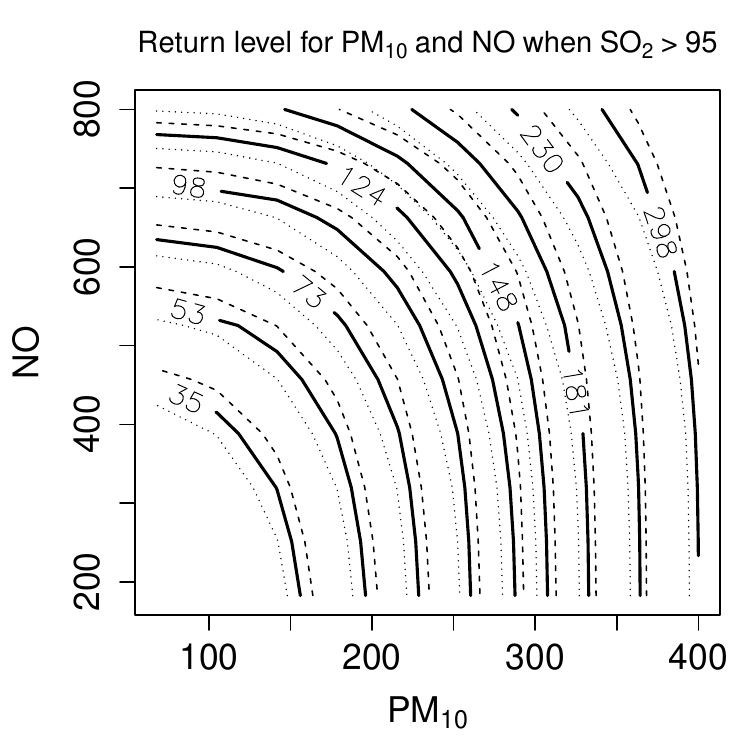} \\
\end{array}
$
\caption{\label{fig:rlevels} Left panel shows the approximate posterior distribution of tail probability with crosses indicating posterior median and $95\%$ (pointwise) credibility interval and a dot indicating the posterior mean. Middle and right panels show the posterior mean (solid line) and posterior $95\%$ credibility interval (dashed lines) of univariate and bivariate return levels associated with return period $1/p$, when respectively fixing two and one components.} 
\end{figure}

The purpose of estimating the extremal dependence is to estimate small tail probabilities as in \eqref{eq:tail copula_tail_approx}. Similar to \cite{Cooley2010}, we define the extreme
event $\{\text{PM}_{10} > 95, \text{NO} > 270, \text{SO}_2 > 95\}$ and infer this probability using the routine \code{pExtDep} with \code{method = "Parametric"}, \code{type = "upper"} and specifying the argument \code{model}. Providing a vector to \code{par} yields a point estimate of the tail probabilities while a matrix (e.g., the posterior sample) with rows of parameter values derives an approximate posterior distribution for such probabilities. We run the following lines
where the object \code{est} and the routine \code{transform} are used to transform values to unit Fr\'{e}chet scale following the steps of \citet{heffernan2004}.
\begin{CodeChunk}
\begin{CodeInput}
R> # Function extracting the Generalized Pareto parameter estimates with a 
R> # threshold set at the 70% quantile
R< est.fun <- function(x){ 
+   x <- na.omit(x)
+   unlist(evd::fpot(x, threshold = quantile(x, probs = 0.7))
+       [c("threshold", "estimate")])
+ }	  
                      
R> est <- apply(winterdat, 2, est.fun)

R> # Transforming univariate data to unit Frechet scale
R> transform <- function (x, data, par){
+   data <- na.omit(data)
+   if(x > par[1]){
+      emp.dist <- mean(data <= par[1])
+      dist <- 1 - (1 - emp.dist) * max(0, 1 + par[3] * (x - par[1]) / 
+         par[2])^(-1 / par[3])
+   }else{dist <- mean(data <= x)}
+   return(-1/log(dist))
+ }

R> th <- c(95, 270, 110, NA, 95)
R> Th <- sapply(c(1:3,5), function(x) 
+    transform(th[x], data = winterdat[, x], par = est[, x]))

R> names(Th) <- colnames(winterdat[c(1:3, 5)])
R> xl.PNS <- bquote("P(" * PM[10]  ~ ">" ~ .(th[1]) * ", NO" ~ ">" 
+    ~ .(th[2]) * ", " * SO[2] ~ ">" ~ .(th[5]) * ")")

R> # Compute and plot probabilities of joint exceedances 
R> P.PNS <- pExtDep(q = Th[colnames(PNS)], type = "upper", 
+    method = "Parametric", model = "HR", par = PNS.hr$stored.vals, 
+    xlab = xl.PNS)
\end{CodeInput}
\end{CodeChunk}
The left panel of Figure \ref{fig:rlevels} provides the approximate posterior distribution of the tail probabilities in \eqref{eq:tail copula_tail_approx} for the extreme events. Note that a smoother output could be obtained by considering $3,000$ iterations (\code{Nsim}) and a burn-in period of $1,000$ iterations (\code{Nburn}). 
Formula \eqref{eq:tail copula_tail_approx} can also be used to derive a possible definition of the so-called {\it joint return level}, see \cite{Beranger2015} for details, i.e., the sequence of values $y_{j;p},j\in J\subset\{1,\ldots,d\}$, that satisfies the equation
$$
p=\prob(Y_j>y_{j;p}, X_i>x_i, j\in J, i\in\{1,\ldots,d\}\backslash J),
$$
for a given small probability $p\in (0,1)$, where $x_i$, with $i\in\{1,\ldots,d\}\backslash J$ is a sequence of fixed high thresholds. An example where the plot routine is used on an object of class \code{ExtDep\_Bayes} to infer joint return levels is given by running the following commands. 
%
\begin{CodeChunk}
\begin{CodeInput}
R> Q.fix <- c(NA, Th[c(2, 4)])
R> PM10.range <- seq(from = est[1, 1], to = 400, by = 5)
R> Q.range <- sapply(PM10.range, transform, data = winterdat[, 1], 
+    par = est[, 1])
R> set.seed(1)

R> # Compute and plot probabilities of return levels
R> rl.PM10 <- plot(x = PNS.hr, type = "returns", Q.fix = Q.fix, 
+    Q.range = Q.range, Q.range0 = PM10.range, labels = expression(PM[10]), 
+    main = bquote("Return level for" ~ PM[10]  ~ "when NO" ~ ">" ~ 
+        .(th[2]) ~ "and" ~ SO[2] ~ ">" ~ .(th[5])), 
+    cex.lab = 1.4, cex.axis = 1.4, subsamp = 0.05)
\end{CodeInput}
\end{CodeChunk}
Only a 5\% subset of the posterior is considered (\code{subsamp = 0.05}) for a faster evaluation but, in practice, we do recommend using the full posterior by leaving \code{subsamp} unspecified. 
Returns levels are displayed when \code{object = "returns"}. In addition, the argument \code{Q.fix} is required and should take the form of 
a vector of length equal to the model dimension ($2$ or $3$), where quantile values can be fixed for some components while others (\code{NA}s) are left to vary. The \code{Q.range} argument provides a vector (or matrix) of quantile values on the unit Fr\'{e}chet scale, for those that aren't fixed. If \code{Q.fix} contains a single \code{NA} then \code{Q.range} must be a vector or a single column matrix. \code{Q.range0} provide the same sequences as \code{Q.range} but on the original scale. This plotting procedure relies on the \code{pExtDep} routine.
The middle panel of Figure~\ref{fig:rlevels} reports the univariate joint return level curve for $\textrm{PM}_{10}$ jointly to the extreme event $\{\text{NO} > 270, \text{SO}_2 > 95\}$, corresponding to the return period $1/p$, which has been estimated by the approximate posterior mean and uncertainty is given by the 95\% (pointwise) credibility intervals. Similarly, the right panel of Figure~\ref{fig:rlevels} depicts the posterior mean and $95\%$ credibility intervals of the estimated contour levels of the bivariate return levels for  $(\textrm{PM}_{10}, \textrm{NO})$ jointly to extreme event $\textrm{SO}_2 > 95$. Note that conditional return levels can be obtained by specifying \code{cond=TRUE}, the conditional event being the fixed event.

%
\subsection[Semi- and non-parametric modeling of the extremal dependence]{Semi- and non-parametric modeling of the extremal dependence}\label{subsec:nonparametric_model}

Common objectives when analyzing extreme values include assessment of the dependence level among extremes, e.g., using the dependence structures in \eqref{eq:evcopula}--\eqref{eq:picklands}, simulation of multiple extremes, estimation of small joint tail probability, e.g., \eqref{eq:stable_tail_approx}--\eqref{eq:tail copula_tail_approx}, and estimation of extreme quantile regions. The next sections describe the steps required to accomplish such goals.  

%
\subsubsection[Extremal dependence estimation via Bernstein polynomials]{Extremal dependence estimation via Bernstein polynomials}
\label{sssec:beed}
%

\noindent\textbf{Theory}. We give here a brief summary of a simple estimation method for the extremal dependence with an arbitrary number of componentwise maxima proposed by \citet{Marcon2017a}. We refer to the aforementioned paper for further details. 

The method consists of two steps: 1) a nonparametric pilot estimation of the Pickands dependence function, 2) a regularization of such estimate by projecting it into a polynomial representation in Bernstein form imposing conditions (C1)--(C2). 
Given some i.i.d. random vectors $\bfY_1,\ldots,\bfY_k$ (approximately) distributed as multivariate EV distribution in \eqref{eq:evd_copula}, the first step is achieved by using the madogram-based estimator of the Pickands dependence function. This is given, for all $\bft\in\resimp$, by 
$$
\widehat{A}_k(\bft)=\frac{\widehat{\nu}_k(\bft)+d^{-1}\sum_{j=1}^d\{t_j/(1+t_j)\}}{1-\widehat{\nu}_k(\bft)-d^{-1}\sum_{j=1}^d\{t_j/(1+t_j)\}},
$$
where 
$$
\widehat{\nu}_k(\bft)=\frac{1}{k}\sum_{i=1}^k\left\{\max_{1\leq j\leq d} F_{k,j}^{1/t_j}(Y_{i,j})-\frac{1}{d}\sum_{j=1}^dF_{k,j}^{1/t_j}(Y_{i,j})\right\},
$$
and $F_{k,j}$ denotes the empirical distribution of the $j$th variable.
For the regularization of the pilot estimate, take $\kappa>d$ and let $\Gamma_{\kappa}$ be the set of multi-indices $\bfalpha=(\alpha_1,\ldots,\alpha_d)^\top\in\{0,1,\ldots,\kappa\}^d$ such that $\alpha_1+\ldots+\alpha_d=\kappa$ and $\alpha_d=\kappa-\alpha_1-\cdots-\alpha_{d-1}$, whose cardinality is denoted by $C_\kappa=|\Gamma_\kappa|$. Let
\begin{equation}\label{multi_berstein}
A_{\kappa}(\bft)=\sum_{\bfalpha\in\Gamma_k}\beta_\bfalpha b_\bfalpha(\bft,\kappa),\quad \bft\in \resimp,
\end{equation}
be the Bernstein--B\'ezier polynomial representation of the Pickands dependence function, where for each $\bfalpha\in\Gamma_\kappa$,
\begin{equation}\label{multi_berstein_basis}
b_\bfalpha(\bft,\kappa)=\frac{\kappa!}{\prod_{j=1}^d \alpha_j!}\prod_{j=1}^{d-1}t_j^{\alpha_j}(1-t_1-\cdots-t_{d-1})^{\alpha_d},\quad \bft\in \resimp,
\end{equation}
is the Bernstein polynomial basis function of index $\bfalpha$ and degree $\kappa$. 

Now, let $\pickset$ be the family of functions $f:\resimp\to[1/d,1]$ satisfying conditions (C1)--(C2), and 
$\pickset_\kappa=\{\bft\mapsto\bfb_\kappa(\bft)\bfbeta_\kappa;\bfbeta_\kappa\in[0,1]^{p_\kappa} \text{ such that } \bfR_\kappa\bfbeta_\kappa\geq \bfr_\kappa\}$ be a sequence of families of constrained multivariate Bernstein–B\'ezier polynomials on $\resimp$, where $\bfb_\kappa(\bft)$ is the row vector $(b_\bfalpha(\bft,\kappa),\,\forall\,\bfalpha\in\Gamma_\kappa)$, $\bfbeta_\kappa$ is a column vector, $\bfR_\kappa$ is a suitable $(q\times p_\kappa)$ matrix of full row rank and $\bfr_\kappa$ is a $(q\times 1)$ vector. The constraint $\bfR_\kappa\bfbeta_\kappa\geq \bfr_\kappa$ on the coefficient vector $\bfbeta_\kappa$ guarantees that each member of $\pickset_\kappa$ satisfies (C1)--(C2). A projection estimator of the Pickands dependence function based on the estimator $\widehat{A}_k(\bft)$ is the solution to the optimization problem
$$
\widetilde{A}_{k,\kappa}=\argmin_{f\in\pickset_\kappa}\| \widehat{A}_k - f\|.
$$
For a finite set of points $\{\bft_u:u=1,\ldots,U\}$, with $U=1,2,\ldots$ and $\bft_u\in\resimp$ such a solution is obtained by finding the minimizer $\widehat{\bfbeta}_\kappa$ of the constrained least-squares problem
\begin{align*}
\widehat{\bfbeta}_\kappa
&=\argmin_{\bfbeta_\kappa\in[0,1]^{N_\kappa}:\bfR_\kappa\bfbeta_\kappa\geq\bfr_\kappa}\frac{1}{U}\sum_{u=1}^U 
\left\{\bfb_\kappa(\bft_u)\bfbeta_\kappa-\widehat{A}_k(\bft_u)\right\}^2 \\
&= \left(\bfb^\top_\kappa\bfb_\kappa \right)^{-1} \bfb_\kappa^\top \widehat{A}_k - \left(\bfb^\top_\kappa\bfb_\kappa \right)^{-1}\bfr_\kappa^\top \bflambda,
\end{align*}
where $\bflambda$ is a vector of Lagrange multipliers.\\

\noindent\textbf{Software implementation}. This method is implemented in the routine \code{beed} of \pkg{ExtremalDep} and its usage is demonstrated through the analysis of heavy rainfall in France. Hydrologists are interested in identifying different geographic regions that differ from each other in that there are clusters of weather stations whose data exhibit substantially different levels of extreme dependence. Within a cluster, climate characteristics are expected to be homogeneous,
whereas they can be quite heterogeneous between clusters. The dataset is available through the \code{PrecipFrance} object which consists of weekly maxima of hourly rainfall (\code{\$precip}) recorded at $92$ weather stations in France, during the Fall season
between 1993 and 2011, yielding a sample size of $k=228$ observations. Coordinates of each station are stored in the list elements \code{\$lat} and \code{\$lon}. Note that hourly rainfall has been appropriately pre-processed and with them the weekly maxima have been already computed from \citet{bernard2013clustering}. 

Through the \code{PAMfmado} routine, the following code chunk applies the algorithm proposed by \citet{bernard2013clustering} to the weekly maxima of hourly rainfall such that the weather stations are divided into seven clusters.
%
\begin{CodeChunk}
\begin{CodeInput}
R> data(PrecipFrance, package = "ExtremalDep")
R> attach(PrecipFrance)

R> nclust <- 7 # number of clusters
R> PAMmado <- PAMfmado(precip, nclust) # Apply clustering algorithm to data
\end{CodeInput}
\end{CodeChunk}
For each cluster, five stations are randomly selected in order to have equal size ($d=5$) clusters and for each group, the Pickands dependence function is estimated using the Bernstein projection estimator based on the madogram with polynomial degree equal $k=7$. To summarise, an estimate of the extremal coefficient is computed using \eqref{eq:extremal_coefficient}, i.e. $\widehat{\varrho}_k=5\widetilde{A}_{k,\kappa}(1/5,\ldots,1/5)$, by running the following commands 
\begin{CodeChunk}
\begin{CodeInput}
R> clust <- PAMmado$clustering
R> d <- 5 # dimension = number of stations
R> stationsn <- matrix(NA, nclust, d)
R> xx <- simplex(d = d, n = 15)
R> fit <- list(length = nclust)
R> est <- vector(length = nclust)
R> set.seed(1)
R> for(i in 1:nclust){
+   # Randomly select 5 stations from cluster i
+   stationsn[i, ] <- sample(which(clust == i), 5)  
+   data_tmp <- precip[, stationsn[i, ]]
+   # Transform margins to unit Frechet scale
+   data_uf_tmp <- trans2UFrechet(data_tmp, type = "Empirical") 
+   # Estimate Pickands dependence function
+   fit[[i]] <- beed(data_uf_tmp, xx, d, "md", "emp", k = 7)
+   # Extract extremal coefficient
+   est[i] <- fit[[i]]$extind}
\end{CodeInput}
\end{CodeChunk}

\begin{figure}[t!]
\centering
\includegraphics[width=0.32\textwidth]{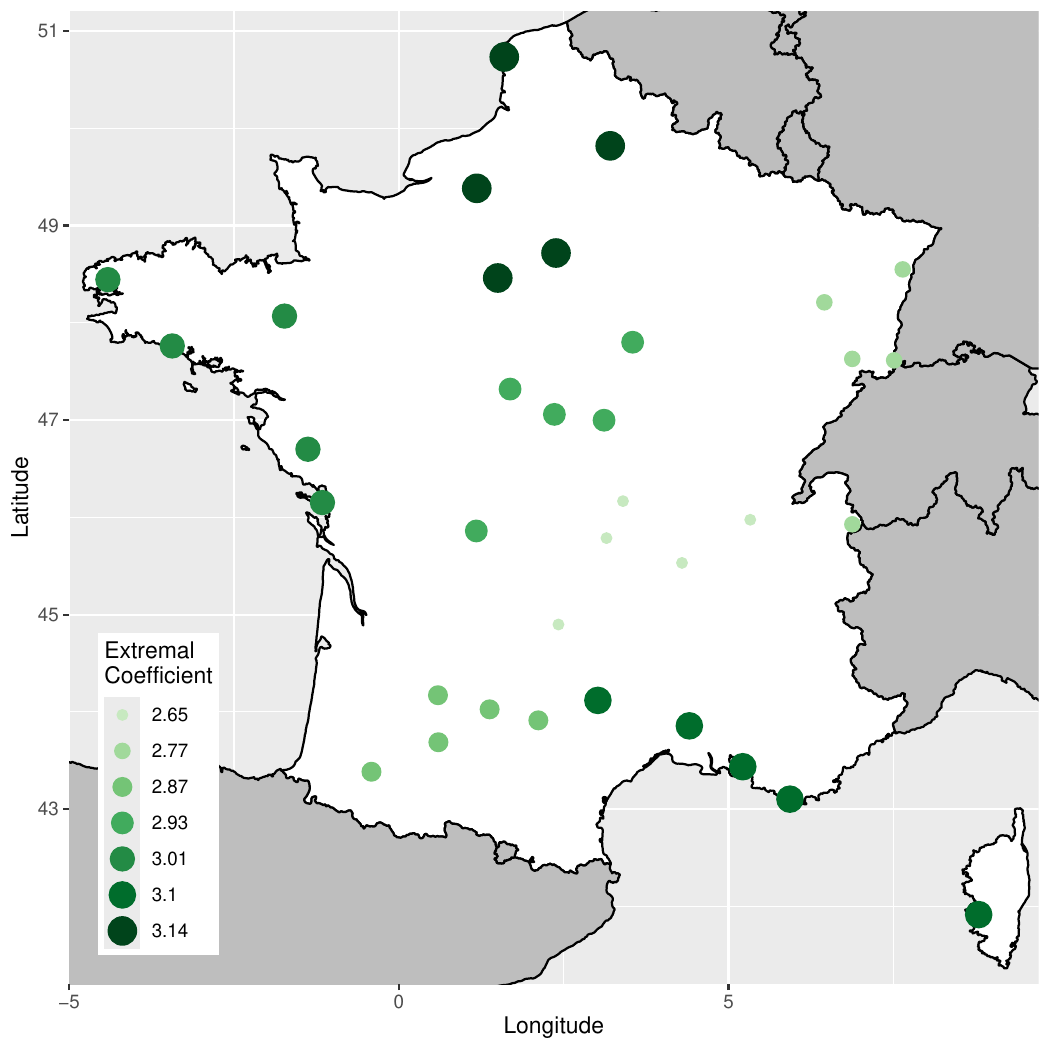} 
\includegraphics[width=0.32\textwidth]{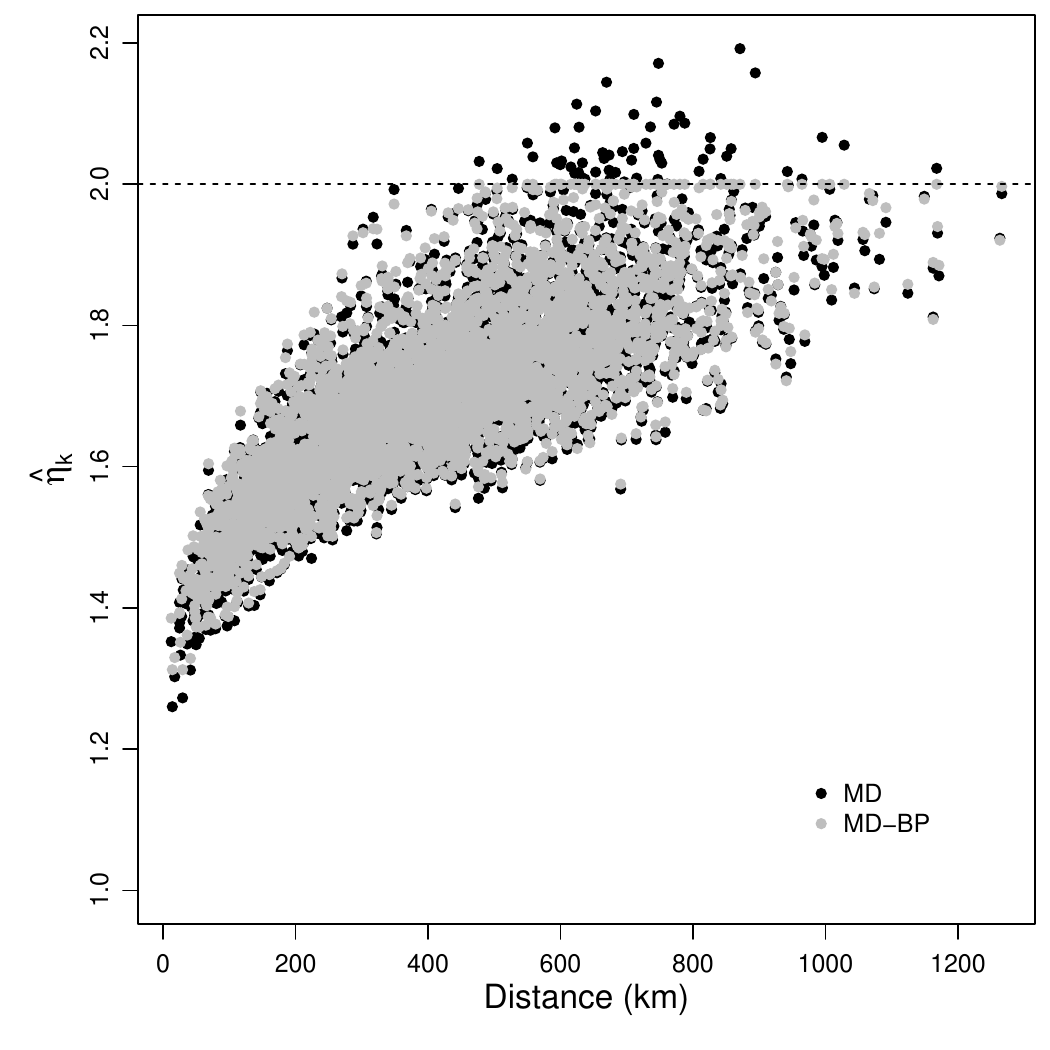}
\caption{\label{fig:french_maxima} Clusters of 35 weather stations and their estimated extremal coefficients in dimension $d = 5$ using French weekly precipitation maxima (left). Evolution of pairwise extremal coefficients with respect to the distance (right).}
\end{figure}

The left panel of Figure \ref{fig:french_maxima} indicates the extremal dependence is strongest in the center of the country, away from the coasts, where the conjunction of different densities of air masses produces extreme rain storms. This is consistent with climatologists expectations since extreme precipitation that affects the Mediterranean coast in the fall is caused by the interaction of southern and mountain winds coming from the Pyr\'en\'ees, C\'evennes and Alps regions. In the north, heavy rainfall is produced by mid-latitude perturbations in Brittany (or regions further north) and Paris. Within clusters, extremes are strongly dependent.
The right panel of Figure~\ref{fig:french_maxima} shows the pairwise extremal coefficients from all $92$ stations, computed through the estimated Pickands dependence functions using the raw madogram estimator (MD) and its Bernstein projection (MD--BP), versus the geodesic distance between sites. We have $\widehat{\varrho}_k \leq 1.5$ for the locations that are less than 200 km apart, meaning that extremes are either strongly or mildly dependent, while for sites more than 200 km apart, we have $\widehat{\varrho}_k > 1.5$, meaning that extremes are at most weakly dependent or even independent. The graph also shows the benefits of the projection method: after projection, the extremal coefficients fall within the admissible range $[1, 2]$.
The \code{beed} function is used for the MD--PB estimator whereas \code{madrogram} implements the raw madogram estimator MD. 
The following set of commands were used for the estimation.
\begin{CodeChunk}
\begin{CodeInput}
R> library(geosphere)

R> pairs <- t(combn(92, 2))   
R> pairs.LLO <- cbind(pairs, lon[pairs[, 1]], lat[pairs[, 1]], 
+    lon[pairs[, 2]], lat[pairs[, 2]])
R> pairs.dist <- apply(pairs.LLO, 1, function(x)  distm(x = x[3:4],  
+    y = x[5:6], fun = distGeo)) / 1000

R> pairs.EC <- pairs.ECBP <- vector(length = nrow(pairs))
R> S <- simplex(d = 2, n = 49)
R> for(i in 1:nrow(pairs)){
+   # Transform margins to unit Frechet scale
+    data.tmp <- trans2UFrechet(precip[,pairs[i, ]], type = "Empirical")
+   # Estimate Pickands dependence function
+    fit.tmp <- beed(data.tmp, S, 2, "md", "emp", k = 7) 
+   # Extract extremal coefficient
+    pairs.ECBP[i] <- fit.tmp$extind
+    pairs.EC[i] <- 2 * madogram(S, data.tmp, "emp")[which(S[, 1] == 0.5)]}
\end{CodeInput}
\end{CodeChunk}
%

%
\subsubsection[Bernstein polynomials modeling and Bayesian nonparametric inference]{Bernstein polynomials modeling and Bayesian nonparametric inference}
\label{sssec:beed_np}
%

\noindent\textbf{Theory}. In many applications, it is crucial to estimate joint probabilities of events such as $p=\prob(Y_1>y_1, Y_2>y_2)$, 
for pair of extreme values value $(y_1,y_2)$. These in turn allow for the estimation of related quantities such as small conditional probabilities $\prob(Y_j>y_j\mid Y_i>y_i)$ with $i,j=1,2$.
Section \ref{sec:back_ground} describes a theoretical framework that can be used to approximate such probabilities and we now briefly describe a Bayesian nonparametric inference method based on Bernstein polynomials introduced by \cite{marcon2016}. 

Let  $\bfy_1,\ldots,\bfy_n$ be $n$ (independent) observations (approximately) distributed as a bivariate GEV distribution. First, a prior distribution on the parameters $\bftheta_j=(\mu_j,\sigma_j,\gamma_j)^\top, j=1,2$, of the marginal GEV distributions is specified as $\Pi(\bftheta_j)=\Pi(\mu_j)\Pi(\sigma_j)\Pi(\gamma_j)\propto 1/\sigma_j$ with $\sigma_j>0$, i.e., a product of 
uniform prior distributions on the real line for $\mu_j$, $\log(\sigma_j)$ and $\gamma_j$. Note that this improper prior distribution leads to a proper posterior distribution \citep{northrop2016}. Samples from the posterior distribution are generated using an adaptive (Gaussian) random-walk Metropolis-Hastings (RWMH) algorithm \citep{haario+st01, garthwaite2016}. For technical details, see Appendix~\ref{app:RWMHalg}.
%
%
%

In the bivariate case, the polynomial Pickands dependence function in Bernstein form becomes
\begin{equation*}\label{eq:bpoly_picka}
 A_{\kappa}(t\mid \bfbeta_\kappa) := \sum_{j=0}^\kappa \beta_j b_j(t; \kappa),
 \qquad t \in\resimp,
\end{equation*}
for $\kappa=0,1,\ldots$, where $\bfbeta_\kappa=(\beta_0,\ldots,\beta_\kappa)^\top$ and the polynomial basis in \eqref{multi_berstein_basis} reduces to 
$$
b_j(t; \kappa) =\frac{\kappa!}{j! (\kappa-j)!}\,j^{j}(1-j)^{\kappa-j}, \quad j=0,\ldots,\kappa.
$$
For fixed degree $\kappa$, \cite{marcon2016} derived the restrictions on $\bfbeta_\kappa$ so that $A_{\kappa}$ satisfies conditions (C2)--(C3) and is therefore a valid Pickands dependence function. They also demonstrated that the polynomial $A_\kappa$ implies that the distribution of the corresponding angular measure can be written as a Bernstein polynomial of degree $\kappa-1$,
\begin{equation}\label{eq:bpoly_angdist}
 H_{\kappa-1}([0,w]\mid \bfeta_\kappa) := 
 \left\{
\begin{tabular}{lcl}
$\sum_{j\leq \kappa-1} \eta_j \;  b_j(w; \kappa-1)$ &  if  & $w\in[0,1)$,\\
1 & if & $w=1$,\\ 
 \end{tabular}
 \right.
\end{equation}
where $\bfeta_\kappa=(\eta_0,\ldots,\eta_{\kappa-1})^\top$.
Additionally,  \cite{marcon2016}  established the conditions on the coefficients $\bfeta_\kappa$ so that $ H_{\kappa-1}$ is the distribution of a valid angular measure and proposed a Bayesian nonparametric procedure for the inference of both $A_{\kappa}$ and $H_{\kappa-1}$, which can then be used to compute an approximation of joint tail probabilities. The proposed method consists of three main steps: 1) the specification of a prior distribution for the polynomial order and coefficients $(\kappa, \bfeta_\kappa)$, 2) the derivation of the likelihood function, 3) the definition of a Markov Chain Monte Carlo (MCMC) algorithm for the posterior distribution computation. In particular, $\Pi(\kappa, \bfeta_\kappa)=\Pi(\bfeta_\kappa\mid\kappa)\,\Pi(\kappa)$, where $\Pi(\kappa)$ is a prior on the polynomial order (e.g., Poisson, negative Binomial, etc.) and $\Pi(\bfeta_\kappa\mid\kappa)=\Pi(\eta_{1},\ldots,\eta_{\kappa-2}\mid p_1,p_0,\kappa)\, \Pi(p_1\mid \kappa,p_0)\,\Pi(p_0)$ is a prior on the coefficients $\bfeta_\kappa$,  where $\Pi(p_1\mid \kappa,p_0)$ and $\Pi(p_0)$ are the priors on the coefficients representative of the atoms $\eta_0=p_0$ and $\eta_{\kappa-1}=1-p_1$ at the edges of the interval $[0,1]$. Such priors are specified as uniform distributions on suitable intervals, which have been chosen to ensure that the resulting Bernstein polynomial satisfies the constraint (C3). Specification of the prior $\Pi(\bfeta_\kappa\mid\kappa)$ induces also a prior on the coefficients  $\bfbeta_\kappa$ of the corresponding  polynomial $A_\kappa(t\mid \bfbeta_\kappa)$, which automatically satisfy constraints (C1)--(C2). To deal with the fact at each MCMC iteration the dimension of $\bftheta_t$ changes with $\kappa$, a trans-dimensional MCMC scheme is considered following \cite{marcon2016}. At iteration $s$, $(\bfeta_{\kappa^{(s)}}^{(s)}, \kappa^{(s)})$ is updated using the proposal distribution $q(\bfeta_\kappa, \kappa \mid \bfeta_{\kappa^{(s)}}^{(s)}, \kappa^{(s)})
=  \Pi(\bfeta_\kappa \mid \kappa)q_\kappa(\kappa \mid \kappa^{(s)})$, where $q_\kappa(\kappa\mid \kappa^{(s)})$ is defined such that if $\kappa^{(s)}=3$, it places mass on $\kappa=4$ with probability $1$ and if $\kappa^{(s)}>3$ it places mass on $\kappa^{(s)} - 1$ and $\kappa^{(s)} + 1$ with equal probability. Finally, the likelihood function is defined as $\mathcal{L}(\bfvartheta)=\prod_{i=1}^m \mathcal{L}(\bfy_i\mid \bfvartheta)$, where $\bfvartheta=(\bftheta_1^\top, \bftheta_2^\top, \bfbeta_\kappa^\top, \kappa)^\top$ and for any $y_j$ such that $(1+\gamma_j(y_j-\mu_j)/\sigma_j)>0$ with $j=1,2$,
%

\begin{align*}
\mathcal{L}(\bfy\mid \bfvartheta)&= G_{\bftheta} \left\{\bfy\mid A_\kappa(t\mid \bfbeta_\kappa) \right\} 
\frac{1}{\sigma_1\sigma_2}
\left(1+\gamma_1\frac{y_1-\mu_1}{\sigma_1}\right)^{-1/\gamma_1-1}
\left(1+\gamma_2\frac{y_2-\mu_2}{\sigma_2}\right)^{-1/\gamma_2-1}
\\
& \times
\left[ \left\{
A_\kappa \left(t\mid\bfbeta_\kappa \right) - tA'_\kappa \left(t\mid\bfbeta_\kappa\right)
\right\}
\left\{
A_\kappa \left(t\mid\bfbeta_\kappa \right) - (1-t)A'_\kappa \left(t\mid\bfbeta_\kappa \right)
\right\} 
+ 
\frac{t(1-t)}{r}A_\kappa^{''}(t\mid\bfbeta_\kappa)
\right] 
\end{align*}
and where $\bftheta=(\bftheta_1^\top, \bftheta_2^\top)^\top$, $t=z_2/r$, $r=z_1+z_2$, 
see \cite{marcon2016} and \cite{beranger2021} for details. The MCMC scheme for the joint inference of marginal distribution and extremal dependence is reported in Algorithm \ref{alg:algo_joint} of Appendix~\ref{app:MCMCalg}. 
For pairs $(y_1^\star, y_2^\star)$ of future unobserved yet extremes values, the joint exceeding probability can be estimated through the Bayesian paradigm using the posterior predictive distribution which can be approximated, given a sample $\bfvartheta_i$ with $i=1,\ldots,M$, from the posterior distribution as
\begin{align}\label{eq_simu_exc_bern}
\prob & (Z_1>z^*_1, Z_2>z^*_2)
\approx  2\sum_{i=1}^M\frac{1}{\kappa_i}\sum_{j=0}^{\kappa_i-2}(\eta_{i,j+1}-\eta_{i,j}) \\ \nonumber
& \times \left\{\frac{(j+1) \, \mathrm{B}\left( \frac{z^{\star}_1}{z^{\star}_1+z^{\star}_2} \; \mid \; j+2,\kappa_i-j-1\right)}{z^{\star}_1} 
 + \frac{(k_i-j-1) \, \mathrm{B}\left( \frac{z^{\star}_2}{z^{\star}_1+z^{\star}_2} \; \mid\; \kappa_i-j,j+1\right)}{z^{\star}_2}
\right\},
\end{align}
where $(Z_1,Z_2)$ are distributed as a bivariate GEV with unit-Fr\'echet margins, $z_j^\star=\sum_{i=1}^M \{1+\gamma_{i,j}(y_j^\star-\mu_{i,j})/\sigma_{i,j}\}^{1/\gamma_{i,j}}$ with $j=1,2$ and $\mathrm{B}(x\mid a,b)$, $x\in[0,1]$, denotes the distribution function of a beta random variable with shape parameters $a,b>0$ \citep[see][for details]{marcon2016}.\\

\noindent\textbf{Application}. We show the utility of the methodology by analyzing the joint extremal behavior of log-returns of exchange rates between Great British Pound and U.S. Dollar (GBP/USD), and Great British Pound and Japanese Yen (GBP/JPY). 
The data are available as \code{logReturns} and consists of the monthly maxima of daily log-returns exchange rates from March 1991 to December 2014 ($286$ observations). Exchange rates are \code{\$USD} and \code{\$JPY}, while \code{\$date\_USD} and \code{\$date\_JPY} are the date when the monthly maxima was attained. 
The following command lines load the data and remove the trend and seasonality from each maxima series using the \code{ts()} and \code{stl()} functions from the \pkg{stats} package.
\begin{CodeChunk}
\begin{CodeInput}
R> data(logReturns, package = "ExtremalDep")

R> # Create time-series objects
R> mm_gbp_usd <- ts(logReturns$USD, start = c(1991, 3), end = c(2014, 12), 
+    frequency = 12)
R> mm_gbp_jpy <- ts(logReturns$JPY, start = c(1991,3), end = c(2014, 12), 
+    frequency = 12)

R> # Decompose time series into seasonal, trend and irregular components
R> seas_usd <- stl(mm_gbp_usd, s.window = "period")
R> seas_jpy <- stl(mm_gbp_jpy, s.window = "period")

R> mm_gbp_usd_filt <- mm_gbp_usd - rowSums(seas_usd$time.series[, -3])
R> mm_gbp_jpy_filt <- mm_gbp_jpy - rowSums(seas_jpy$time.series[, -3])
\end{CodeInput}
\end{CodeChunk}
\begin{figure}[t!]
\centering
\includegraphics[width=0.49\textwidth]{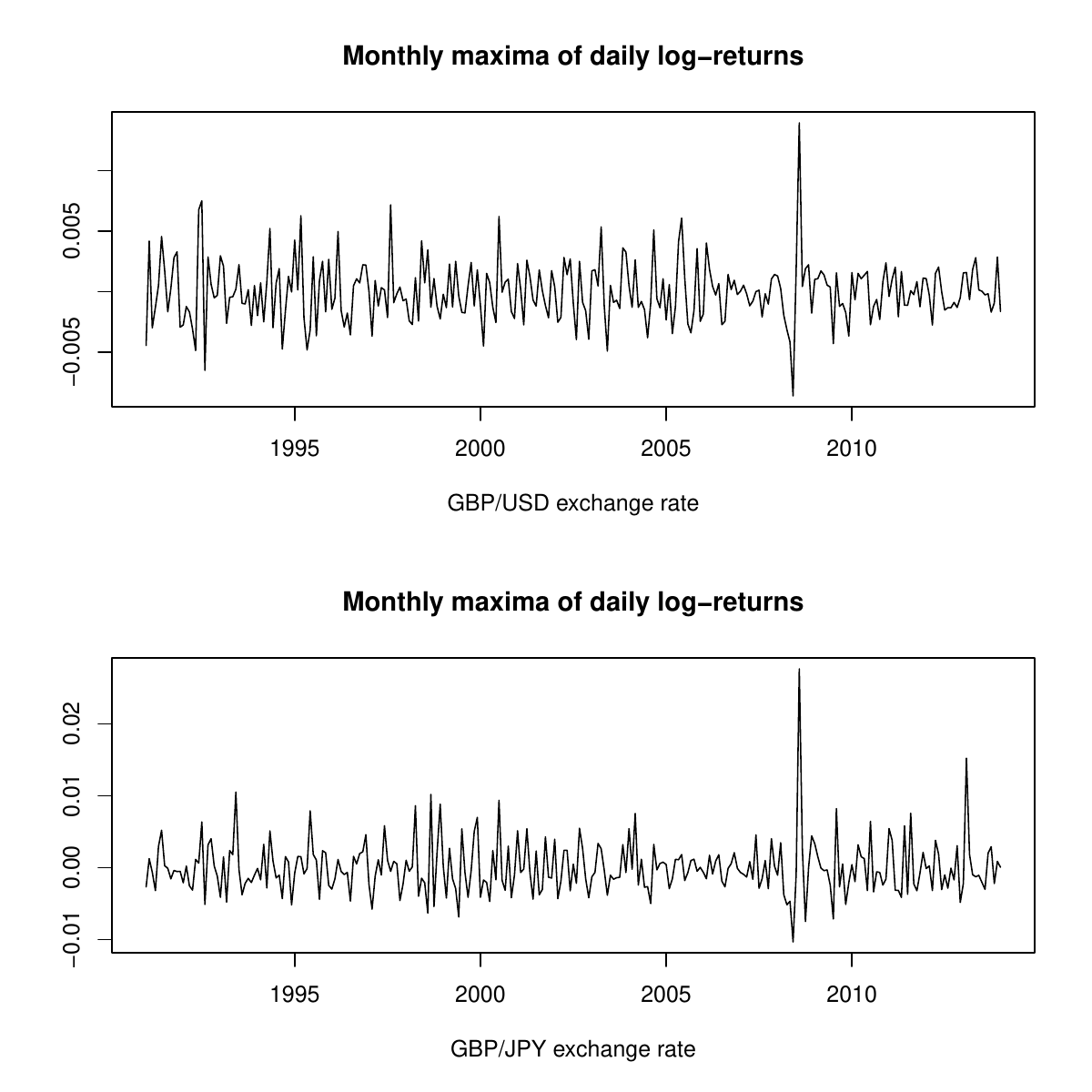} 
\includegraphics[width=0.49\textwidth]{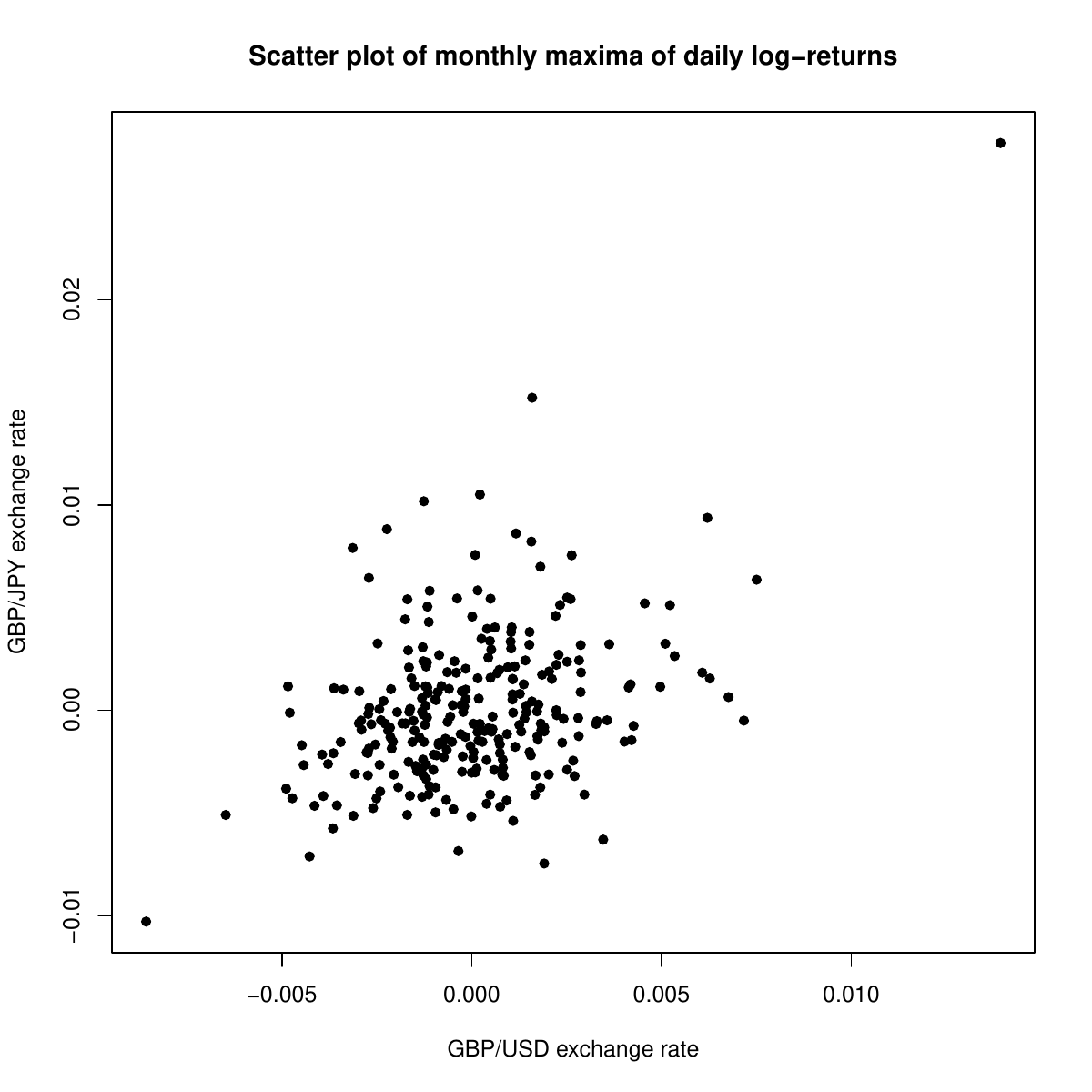} 
\caption{\label{fig:logRet} Detrended and deseasonalised times series of monthly-maxima of log-returns of GBP/USD and GBP/JPY exchange rates (left) and corresponding scatterplot (right).}
\end{figure}
The top-left and bottom-left panels of Figure~\ref{fig:logRet} display the detrended and deseasonalised maxima and the scatterplot in the right panel shows strong dependence between the extremes of both exchanges rates. 

The extremal dependence structure is then estimated using the Bayesian nonparametric framework described above by running the following commands
\begin{CodeChunk}
\begin{CodeInput}
R> hyperparam <- list(mu.nbinom = 3.2, var.nbinom = 4.48)
R> mm_gbp <- cbind(as.vector(mm_gbp_usd_filt), as.vector(mm_gbp_jpy_filt)) * 100

R> set.seed(123)
R> # Estimate dependence structure using Bayesian nonparametric
R> par0 <- c(-0.1, 0.1, -0.1)
R> gbp_mar <- fExtDep.np(x = mm_gbp, method = "Bayesian", par10 = par0, 
+    par20 = par0, sig10 = 0.1, sig20 = 0.1, k0 = 5,  
+    hyperparam = hyperparam, nsim = 5e+4)
\end{CodeInput}
%\begin{CodeOutput}
%Preliminary on margin 1 
%
%|=====================================================================| 100%
%Preliminary on margin 2 
%
%|=====================================================================| 100% 
%
%Estimation of the extremal dependence and margins 
%|=====================================================================| 100%
%\end{CodeOutput}
\begin{CodeInput}
R> diagnostics(gbp_mar)
\end{CodeInput}
\end{CodeChunk}
Note that the data has been rescaled by a factor of $100$ to avoid numerical instabilities due to the small scale of the data.
The \code{fExtDep.np} routine allows for nonparametric estimation of the extremal dependence and \code{method = "Bayesian"} specifies that such a dependence is in Bernstein polynomial form and
a Bayesian approach is used for the inference. The argument \code{mar.fit = TRUE} (default) allows for joint estimation of the margins and dependence while \code{mar.prelim = TRUE} (default) fits the marginal distributions using the RWMH algorithm and the \code{fGEV} routine, in order to obtain starting values for the marginal parameters. The prior distribution for $\kappa$ is set to be a negative binomial on $\kappa - 3$ with mean $3.2$ and variance $4.48$, and for the point masses $p_0$ and $p_1$ uniform distributions on $[0, 0.5]$ and $[a,b]$ respectively, where $a = \max\{ 0, (\kappa-1)p_0-\kappa/2+1\}$ and $b=(p_0+\kappa/2-1)/(\kappa-1)$. Note that both the negative binomial and the Poisson distributions are convenient choices for the prior distribution of the polynomial order $\kappa$ and the selected hyper-parameter values simply allow for a broad range of admissible values. For the prior of the point mass $p_1$ conditionally on $\kappa$ and $p_0$, the bounds of the uniform distribution are set to fulfill the conditions on the coefficients $\bfeta_\kappa$. We refer to \citet{marcon2016} for a comprehensive discussion. In this example, only the hyper-parameters need to be specified since \code{prior.k = "nbinom"} and \code{prior.pm = "unif"} by default. Lastly, a two-column object \code{mm\_gbp} representing the data is created.

\begin{figure}[t!]
\centering
\includegraphics[width=0.49\textwidth]{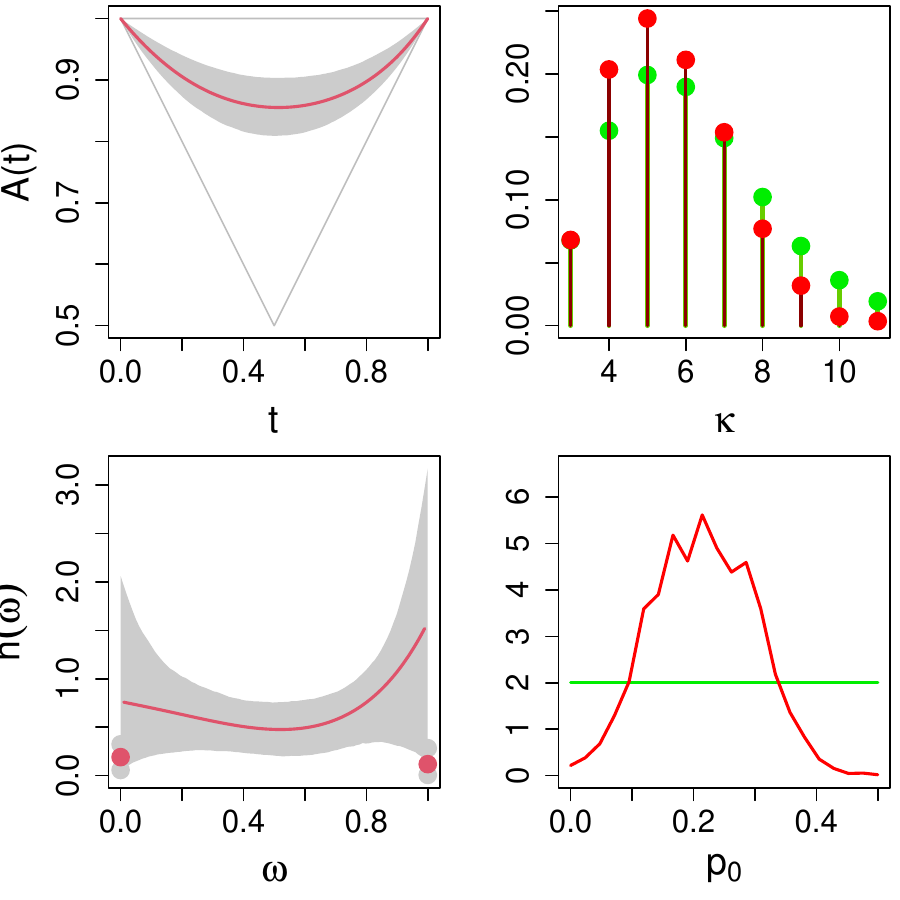}
\includegraphics[width=0.49\textwidth]{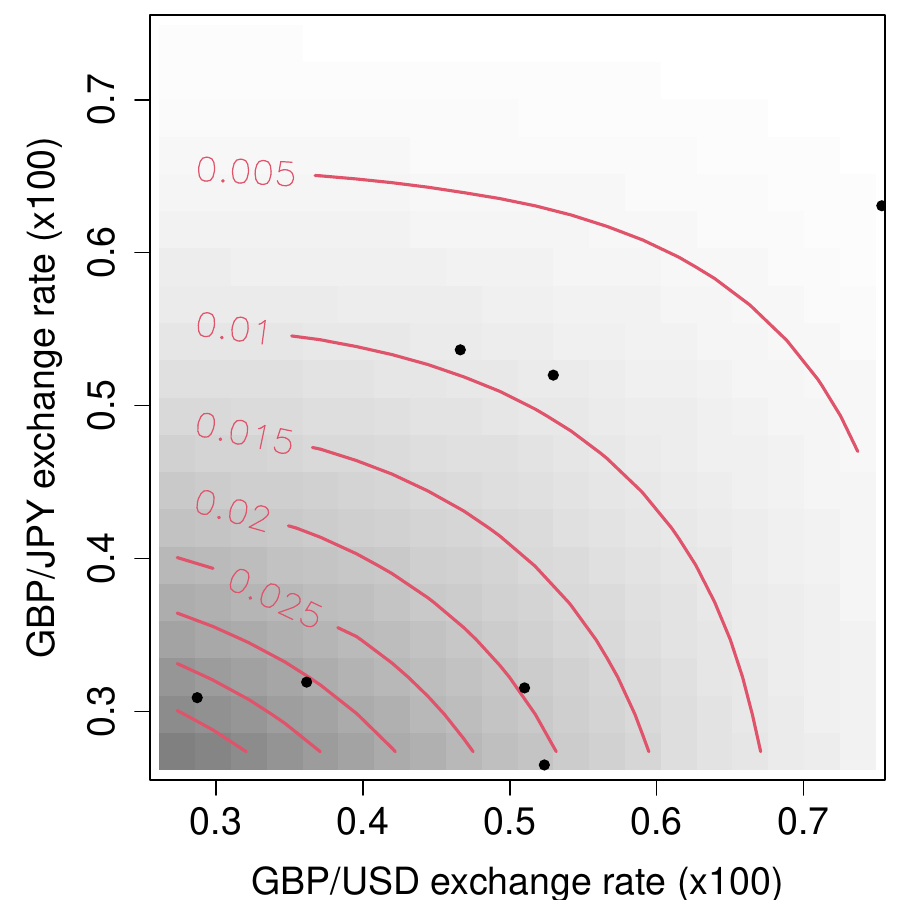}
\caption{\label{fig:sum_logRet} Outputs of the \code{summary.ExtDep} (left) and \code{returns} (right) functions. }
\end{figure}

The \code{gbp\_mar} object is of class \code{ExtDep\_npBayes} and contains the posterior samples for all the parameters: point mass $p_0$ and $p_1$, polynomial coefficients $\boldsymbol{\eta}$ and degree $\kappa$, marginal parameters $\bftheta_1$ and $\bftheta_2$, respectively reported by the arguments \code{pm}, \code{eta}, \code{k}, \code{mar1} and \code{mar2}. Its also contains binary vectors indicating the accepted marginal and dependence proposals (\code{accepted.mar1}, \code{accepted.mar2} and \code{accepted}) and the marginal proposals that were rejected right away for not being in the parameter space (\code{straight.reject1} and \code{straight.reject2}). It also includes acceptance probabilities at each step (\code{acc.vec}, \code{acc.vec.mar1} and \code{acc.vec.mar2}), the marginal scaling parameters $\tau_1$ and $\tau_2$ (\code{sig1.vec} and \code{sig2.vec}), as well as some of the inputs.
\begin{figure}[t!]
\centering
\includegraphics[width=\textwidth]{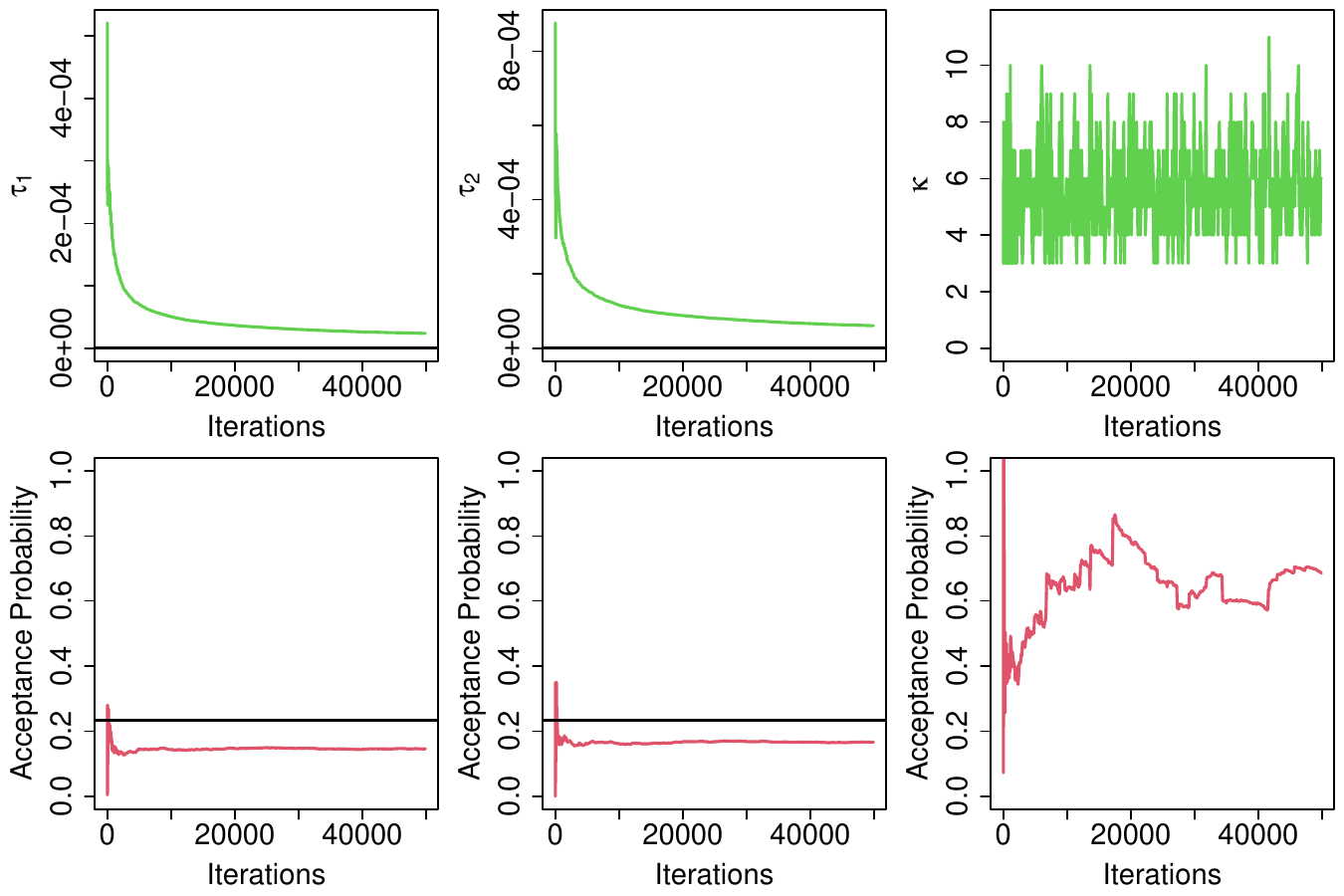}
\caption{\label{fig:diag_logRet} Diagnostic plots for the MCMC algorithm. Left and centre columns focus on the marginal components, illustrating the respective sampler scaling parameter and acceptance probability. Right column focuses on the dependence structure, illustrating the value of the polynomial degree throughout the algorithm and acceptance probability.}
\end{figure}
The \code{diagnostics} function investigates the convergence of the algorithm, producing Figure~\ref{fig:diag_logRet}. The top panels display the scaling parameters $\tau_1$ and $\tau_2$ from the marginal proposals and the polynomial degree $\kappa$, as function of the iterations. The bottom panels show the corresponding acceptance probability with the desired acceptance probability of $0.234$ (horizontal black solid lines). For both margins, the acceptance rate stabilizes at a level slightly below the nominated level of $0.234$ which yields decreasing values of the corresponding $\tau$ parameters. One could be tempted to increasing the number of MCMC iterations or the burn-in period but given the range of values taken, this would have little impact. In addition, the first two rows of the Figure~\ref{fig:diag_logRet} indicate that acceptance probability for the marginal estimates is slightly below the nominated levels. Actual convergence of the marginal chains can be confirmed using traceplots on the columns of the objects \texttt{gbp\_mar\$mar1} and \texttt{gbp\_mar\$mar2}. Overall, Figure~\ref{fig:diag_logRet} suggests a burn-in period of $30,000$ iterations and we now run the following commands
%
%
\begin{CodeChunk}
\begin{CodeInput}
R> # Compute summary statistics of the MCMC output
R> gbp_mar_sum <- summary(object = gbp_mar, burn = 30000, plot = TRUE)
					
R> mm_gbp_rg <- apply(mm_gbp, 2, quantile, c(0.9, 0.995))

R> y_gbp_usd <- seq(from = mm_gbp_rg[1, 1], to = mm_gbp_rg[2, 1], length = 20)
R> y_gbp_jpy <- seq(from = mm_gbp_rg[1, 2], to = mm_gbp_rg[2, 2], length = 20)
R> y <- as.matrix(expand.grid(y_gbp_usd, y_gbp_jpy, KEEP.OUT.ATTRS = FALSE))

R> # Compute and plot return levels
R> ret_marg <- returns(x = gbp_mar, summary.mcmc = gbp_mar_sum, y = y, 
+    plot = TRUE, data = mm_gbp, xlab = "GBP/USD exchange rate (x100)", 
+    ylab = "GBP/JPY exchange rate (x100)")				
\end{CodeInput}
\end{CodeChunk} 
The \code{summary} routine applied to objects of class \code{ExtDep\_npBayes} computes summary statistics of the MCMC output, including posterior sample, mean and $95\%$ credibility intervals (the latter can be modified using argument \code{cred}) for the angular density and Pickands dependence function. Posterior mean and credibility intervals are also computed for all parameters. Setting \code{plot=TRUE} displays the posterior mean and credibility intervals for both angular density and Pickands dependence function, as well as the prior and posterior distribution for the point mass $p_0$ and the polynomial degree $\kappa$. 
The \code{returns} routine inputs objects of class \code{ExtDep\_npBayes} and outputs of the \code{summary} function (via the arguments \code{x} and \code{summary.mcmc}) to compute exceeding probabilities as defined in \eqref{eq_simu_exc_bern}, for extreme values  specified by \code{y}. The argument \code{plot} calls the \code{plot} routine on the object \code{x} to visualize such probabilities as long as \code{y} defines a square grid and \code{data} adds the relevant datapoints. Usage of the \code{summary} and \code{returns} functions is provided below with graphical outputs presented in Figure~\ref{fig:sum_logRet}.
As mentioned at the beginning of this section, computing small conditional probabilities such as ${\prob(\textrm{GBP/USD}>q_1\mid\textrm{GBP/JPY}>q_2)}$ or ${\prob(\textrm{GBP/JPY}>q_2\mid\textrm{GBP/USD}>q_1)}$ can be of interest. We proceed by taking $(q_1,q_2)\approx(0.0069, 0.0102)$ which corresponds to the observed marginal $99\%$ quantiles. The joint probability of exceedance is computed (through \code{returns}) using the approximation $\prob(\textrm{GBP/USD}>q_1, \textrm{GBP/JPY}>q_2) \approx \prob(Z_1>z_1^*, Z_2>z_2^*) \approx 0.0044$, where $z_1^*$ and $z_2^*$ are the transformation of $q_1$ and $q_2$ to unit Fr\'{e}chet scale. Using the \code{pGEV} function to evaluate the marginal probability of exceedance, we obtain $\prob(\textrm{GBP/USD}>q_1\mid\textrm{GBP/JPY}>q_2) \approx 0.3169$ and $\prob(\textrm{GBP/JPY}>q_2\mid\textrm{GBP/USD}>q_1) \approx 0.2767$. The computations are presented in 
the code below.
\begin{CodeChunk}
\begin{CodeInput}
R> qs <- apply(mm_gbp, 2, quantile, c(0.99))

R> # Compute joint probability of exceedance
R> jointP <- returns(x = gbp_mar, summary.mcmc = gbp_mar_sum, 
+    y = matrix(qs, ncol = 2)) 

R> # Compute conditional probabilities of exceedance
R> par1 <- gbp_mar_sum$mar1.mean
R> par2 <- gbp_mar_sum$mar2.mean
R> jointP / pGEV(qs[1], loc = par1[1], scale = par1[2], shape = par1[3], 
+    lower.tail = FALSE)
R> jointP / pGEV(qs[2], loc = par2[1], scale = par2[2], shape = par2[3], 
+    lower.tail = FALSE)		
\end{CodeInput}
\end{CodeChunk} 
The model's goodness of fit provided can be verified graphically at the marginal and dependence level. First, marginal density histograms can be plotted together with GEV density curves evaluated at the posterior samples of the marginal parameters. Second, after standardizing the data to unit Fr\'{e}chet marginals, a density histogram of the angular components with the largest radial component can be used as an empirical estimate of the angular density, to which the model estimates (posterior mean, $0.05$ and $0.95$ quantile) can be added.  Figure~\ref{fig:GBP_GoF}, given in Appendix~\ref{app:GoF}, provides graphical evidence that the proposed model does provide a good fit to the data.

%
\subsubsection[Extremes simulation]{Simulation of extreme values}
\label{sssec:sim_ev}
%

\noindent\textbf{Theory}. In environmental statistics, there is active research in the field of ``stochastic weather generators'' which aim at simulating realisations from atmospheric variables according to some stochastic representation. A challenging task is the simulation of multivariate
extremes, since their extremal dependence (Pickands dependence function or angular measure) is an infinite dimensional parameter of the multivariate GEV distribution and is therefore challenging to estimate. The ability to simulate multivariate extremes permits to approximate the tail probabilities in \eqref{eq:stable_tail_approx} and \eqref{eq:tail copula_tail_approx}, even though few extremes are available in the dataset. 

For simplicity, most of the simulation methods assume that the extremal dependence complies with some parameter model, while few attempt to consider a nonparametric approach. \citet{Marcon2017} proposed a flexible procedure for sampling from bivariate extremes with a semiparametric dependence structure, which is summarized as follows. 
For every small probabilities $p_1$ and $p_2$ such that $L(p_1,p_2)\in[0,1]$ and $R(p_1,p_2)\in[0,1]$, the tail probabilities in \eqref{eq:stable_tail_approx} and \eqref{eq:tail copula_tail_approx} can be approximated by $\prob\{X_1>Q_1(p_1)\, \text{or} \, X_2>Q_2(p_2)\}\approx L(p_1,p_2)$ and $\prob\{X_1>Q_1(p_1)\, \text{and} \, X_2>Q_2(p_2)\}\approx R(p_1,p_2)$. 
For such probabilities $p_1$ and $p_2$,  we have  that $L(p_1,p_2)=2\expect\{\max(p_1 W, p_2(1-W)\}$ and $R(p_1,p_2)=2\expect\{\min(p_1 W, p_2(1-W)\}$, where $W$ is a random variable on $[0,1]$ distributed according to the angular distribution $H$. Let $(Z_1,Z_2)=R(W, 1-W)$, where $R$ is a unit-Pareto random variable, then $\prob(Z_1>z_1\, \text{or} \, Z_2>z_2)=2\expect\{\max(W/z_1,(1-W)/z_2)\}$ and $\prob(Z_1>z_1\, \text{and} \, Z_2>z_2)=2\expect\{\min(W/z_1,(1-W)/z_2)\}$.
\citet{Marcon2017} proposed to model $H$ through Bernstein polynomials and demonstrated that $H$ can be written as a finite mixture of beta distributions with weights defined by a suitable transformation of the polynomial coefficients. This leads to a simple algorithm to sample from $H$ (see Algorithm 1) and an algorithm for sampling observations from the tail of a bivariate distribution (see Algorithm 3), which can be used to approximate the corresponding tail probabilities. Briefly, from \eqref{eq:uniGEV} we have that, for sufficiently large $n$ and sufficiently small $p_j=p_{n,j}$ (with $p_j\to 0$ as $n\to\infty$), $Q_j(p_j)\approx \mu_j+\gamma_j\{(np_j)^{-\gamma_j}-1\}/\sigma_j$, for $j=1,2$. Let $u_1$ and $u_2$ be two high thresholds such that, for every $p_1$ and $p_2$ for which the above approximations hold and for which, in addition, $L(p_1,p_2)\in[0,1]$ and $R(p_1,p_2)\in[0,1]$, we have $Q_j(p_j)>u_j$, $j=1,2$. For the failure regions  $\mathcal{A}_\bfu = \left\{ (v_1, v_2): v_1 >u_1 \textrm{ or } v_2>u_2\right\}$ and  $\mathcal{B}_\bfu = \left\{ (v_1, v_2): v_1 >u_1 \textrm{ and } v_2>u_2\right\}$ we have $\prob(\bfX\in \mathcal{A}_\bfu)\approx \prob(Z_1>u^\star_1\, \text{or} \, Z_2>u^\star_2)$ and $\prob(\bfX\in \mathcal{B}_\bfu)\approx \prob(Z_1>u^\star_1\, \text{and} \, Z_2>u^\star_2)$, where $u^\star_j=(1+\gamma_j(u_j-\mu_j)/\sigma_j)^{1/\gamma_j}$, for $j=1,2$. By simulating a large sample of angular components $(w_1,\ldots,w_N)$ from $H$ and radial components  $(r_1,\ldots,r_N)$ from a unit-Pareto distribution, we compute $\bfz_i=2(r_iw_i,r_i(1-w_i))$, for $i=1,\ldots, N$ and estimate of the probability of falling in  $\mathcal{A}_\bfu$ and $\mathcal{B}_\bfu$ by  
\begin{equation}\label{eq:empirical_prob}
\hat{p}_{\mathcal{A}_{\boldsymbol{u}}} = \frac{1}{N} \sum_{i=1}^N \indic\left( z_{i,1} > u^\star_1 \textrm{ or } z_{i,2} >u^\star_2\right), 
\quad 
\hat{p}_{\mathcal{B}_{\boldsymbol{u}}} = \frac{1}{N} \sum_{i=1}^N \indic\left( z_{i,1} > u^\star_1 \textrm{ and } z_{i,2} >u^\star_2\right).
\end{equation}

\noindent\textbf{Application.} The methodology is illustrated on the \code{Parcay-Meslay} dataset which consists of daily maxima of hourly wind speed (\code{WS}) and wind gust (\code{WG}) in meters per second (m/s) and differential of daily range of the hourly air pressure (\code{DP}) at sea level in millibars. Measurements are taken in the city of Par\c{c}ay-Meslay, located in the northwest of France, from July 2004 to July 2013. For this analysis, we focus on positive values of \code{DP} implying an increase in the daily air pressure level. High air pressure levels are associated with a high content of water vapor in the air which often occurs in stormy weather and leads to strong winds. For brevity, we focus on the relationship between \code{WS} and \code{DP}. The GEV parameters are first estimated using the point process approach \citep[e.g.,][Ch.~7]{Coles2001}, implemented in the \code{fpot} routine of the \pkg{evd} package \citep{Stephenson2002}, where the observed $0.9$ marginal quantiles are set as suitable thresholds. The margins are transformed to unit-Fr\'{e}chet scale using the \code{trans2UFrechet} routine with argument \code{type = "empirical"}, meaning the transformation $y_{i,j} = 1 / \left\{ 1 - F_{n,j} (x_{i,j})\right\}, i=1, \ldots, n, j=1,2$ is applied, where $F_{n,j}$ denotes the empirical distribution of the $j$-th component.
These initial steps can be implemented by running the following lines
%
\begin{CodeChunk}
\begin{CodeInput}
R> data(WindSpeedGust, package = "ExtremalDep")
R> years<- format(ParcayMeslay$time, format = "%Y")
R> attach(ParcayMeslay[which(years %in% c(2004:2013)), ])

R> WS_th <- quantile(WS, 0.9)
R> DP_th <- quantile(DP, 0.9)

R> # Peaks over threshold modelling using the point processes representation
R> pars.WS <- evd::fpot(WS, WS_th, model = "pp")$estimate
R> pars.DP <- evd::fpot(DP, DP_th, model = "pp")$estimate

R> # Transform margins to unit Frechet scale
R> data_uf <- trans2UFrechet(cbind(WS, DP), type = "Empirical")
\end{CodeInput}
\end{CodeChunk}

\begin{figure}[t!]
\centering
\includegraphics[width=0.32\textwidth]{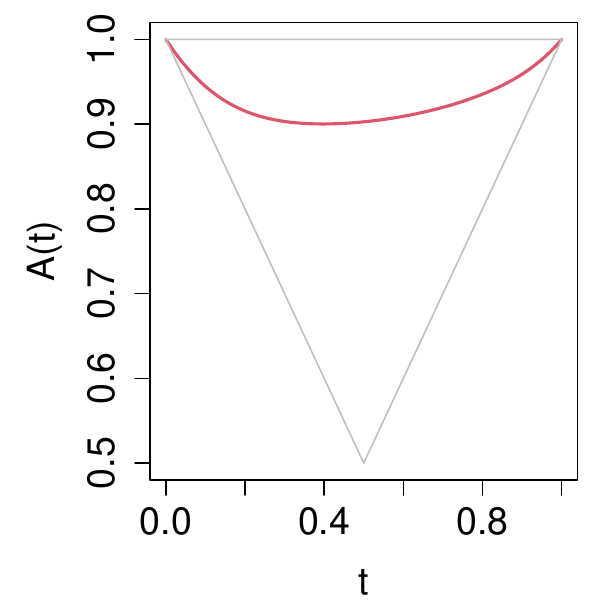}
\includegraphics[width=0.32\textwidth]{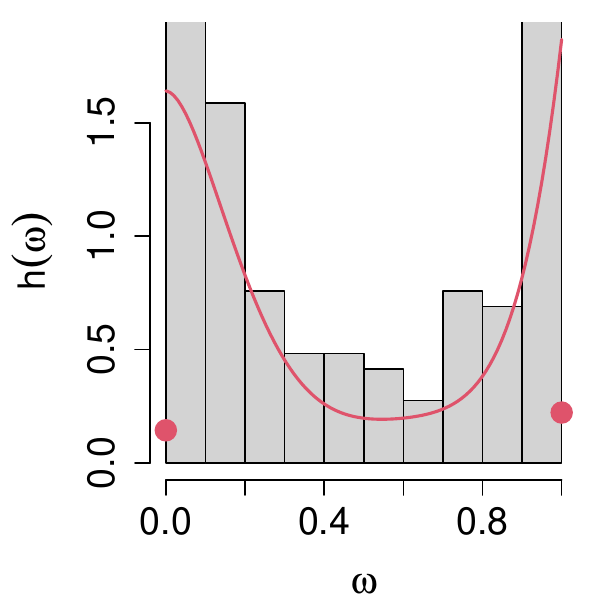}
\includegraphics[width=0.32\textwidth]{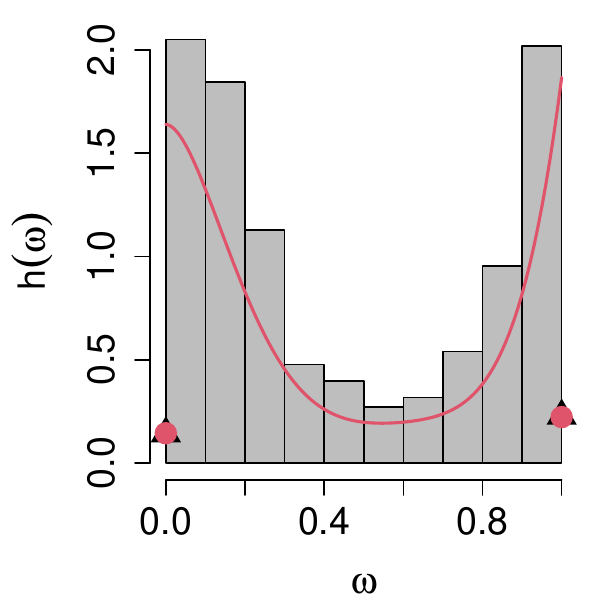}
\caption{\label{fig:winds} Estimated Pickands dependence function (left) and angular density (middle and right). The middle panel displays the histogram of the data while the right panel displays a histogram of the simulated pseudo-angles as well as the point masses (black triangles).}
\end{figure}

The angular distribution is then estimated using the approximate likelihood approach \citep[see][]{ , Beranger2015}. Observations with radius component greater than their $90\%$ quantile are selected and the routine \code{fExtDep.np} is then called with arguments \code{method = "Frequentist"} and \code{type = "maxima"} to maximize the likelihood of a polynomial angular distribution in Bernstein form through the non-linear optimization routine \code{nloptr} from the \pkg{nloptr} package \citep{nloptr} subject to constraints established in \citet{Marcon2017a}.
Empirical studies \citep{Marcon2017a} suggest that the polynomial degree $\kappa=10$ is enough to represent flexible dependence structures. When \code{type = "rawdata}, data are extracted using a threshold on the radial component set by the argument \code{u} and the likelihood for a sample maxima written as function of the Pickands dependence function in Bernstein form is optimized. In addition, when \code{mar.fit = TRUE} then marginal empirical transformation to unit Fr\'{e}chet of the data is applied. The \code{plot} routine with \code{type = "summary"} displays graphical summaries (Pickands dependence function and angular density) of the estimated dependence structure (see left and middle panels of Figure~\ref{fig:winds}). A moderate level of dependence is observed as well as point masses at the vertices. Given that the middle panel of Figure~\ref{fig:winds} combines empirical information with a summary of the fitted model, it can be leveraged to highlight the good fit of the model. The routine \code{rExtDep} with arguments \code{model = "semi.bvevd"} and \code{angular = TRUE} generates pseudo-angles according to Algorithm 1 of \citet{Marcon2017}. The right panel of Figure~\ref{fig:winds} presents a histogram of $1,000$ randomly generated angles and point-masses (black triangles), the red line and dots represent the estimated dependence structure. The corresponding code is 
%
\begin{CodeChunk}
\begin{CodeInput}
R> rdata <- rowSums(data_uf) # Computes radial component
R> r0 <- quantile(rdata, probs = 0.9)
R> extdata <- data_uf[rdata >= r0, ] # Select data with large radial component

R> # Estimate angular density
R> SP_mle <- fExtDep.np(x = extdata, method = "Frequentist", k0 = 10, 
+    type = "maxima")
R> plot(x = SP_mle, type = "summary")

R> # Generate angular data
R> SP_wsim <- rExtDep(n = 1000, model = "semi.bvevd",  
+    param = SP_mle$Ahat$beta, angular = TRUE)
\end{CodeInput}
\end{CodeChunk}

\begin{figure}[t!]
\centering
\includegraphics[width=0.5\textwidth]{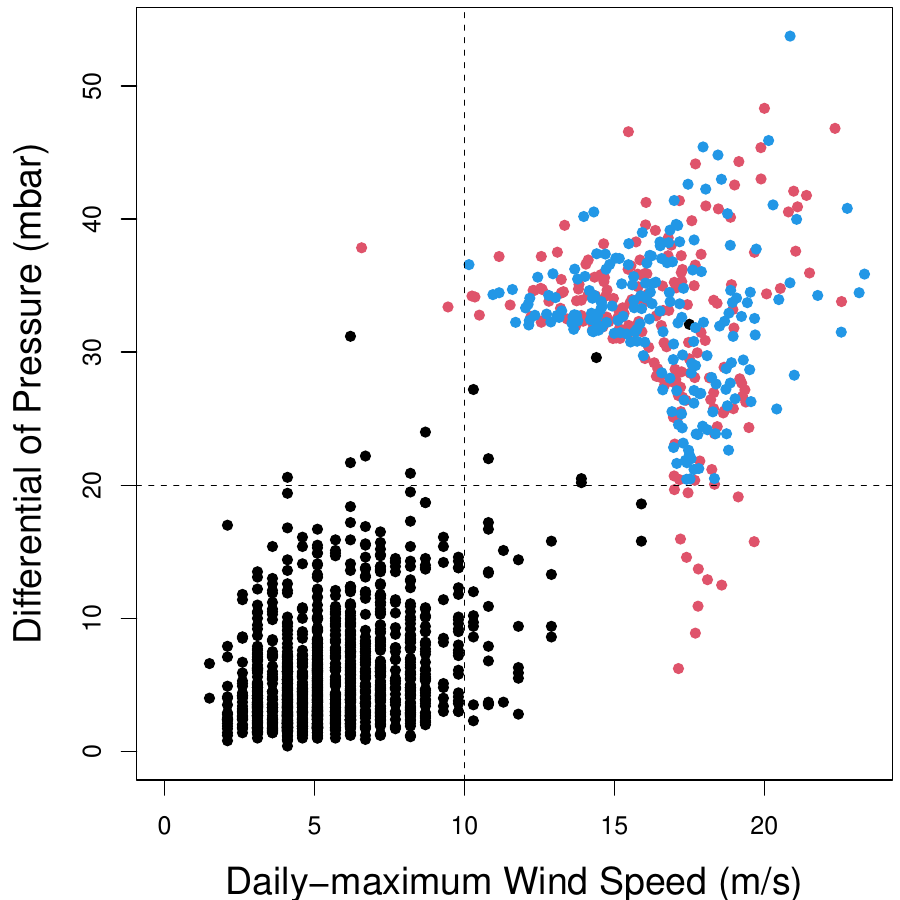}
\caption{\label{fig:failure} Observed (black dots) differential of pressure (mbar) and daily-maximum wind speed (m/s) in Par\c{c}ay-Meslay, France between 2004 and 2013. Simulation of 200 observations from the failure regions $\mathcal{A}_{\boldsymbol{u}}$ (red dots) and $\mathcal{B}_{\boldsymbol{u}}$ (blue dots) with $\boldsymbol{u}=(10,20)$ (dashed lines).}
\end{figure}

The routine \code{rExtDep} performs random generation of bivariate maxima (\code{model = "semi.bvevd"}) and bivariate exceedances  (\code{model = "semi.bvexceed"}), according to Algorithm 2 and 3 of \citet{Marcon2017}. When \code{angular = TRUE}, only angular components are generated, no matter the \code{model} argument, meaning that setting \code{model = "semi.bvexceed"} in the above call of \code{rExtDep} would produce the same result. The argument \code{mar} allows for transformations to GEV margins. When \code{model = "semi.bvexceed"}, one can choose to simulate bivariate observations from the failure regions $\mathcal{A}_{\bfu}$ (\code{exceed.type = "or"}) or $\mathcal{B}_{\bfu}$ (\code{exceed.type = "and"}), where $\bfu$ is a suitable threshold specified by \code{threshold}.  The command lines below generate
$200$ observations from both failure regions with threshold $\bfu=(10,20)$, above the marginal $90\%$ quantiles (see Figure~\ref{fig:failure} for an illustration).
\begin{CodeChunk}
\begin{CodeInput}
R> set.seed(10)

R> # Generate exceedances in at least one component
R> SP_exceed_or <- rExtDep(n = 200, model = "semi.bvexceed", 
+    param = SP_mle$Ahat$beta, mar = rbind(pars.WS, pars.DP),  
+    threshold = c(10, 20), exceed.type = "or")

R> # Generate exceedances in both components
R> SP_exceed_and <- rExtDep(n = 200, model = "semi.bvexceed", 
+    param = SP_mle$Ahat$beta, mar = rbind(pars.WS, pars.DP),  
+    threshold = c(10, 20), exceed.type = "and")
\end{CodeInput}
\end{CodeChunk}

\begin{figure}[t!]
\centering
\includegraphics[width=0.4\textwidth, page=2]{WindGust_pFailure_v2.pdf}
\includegraphics[width=0.4\textwidth, page=1]{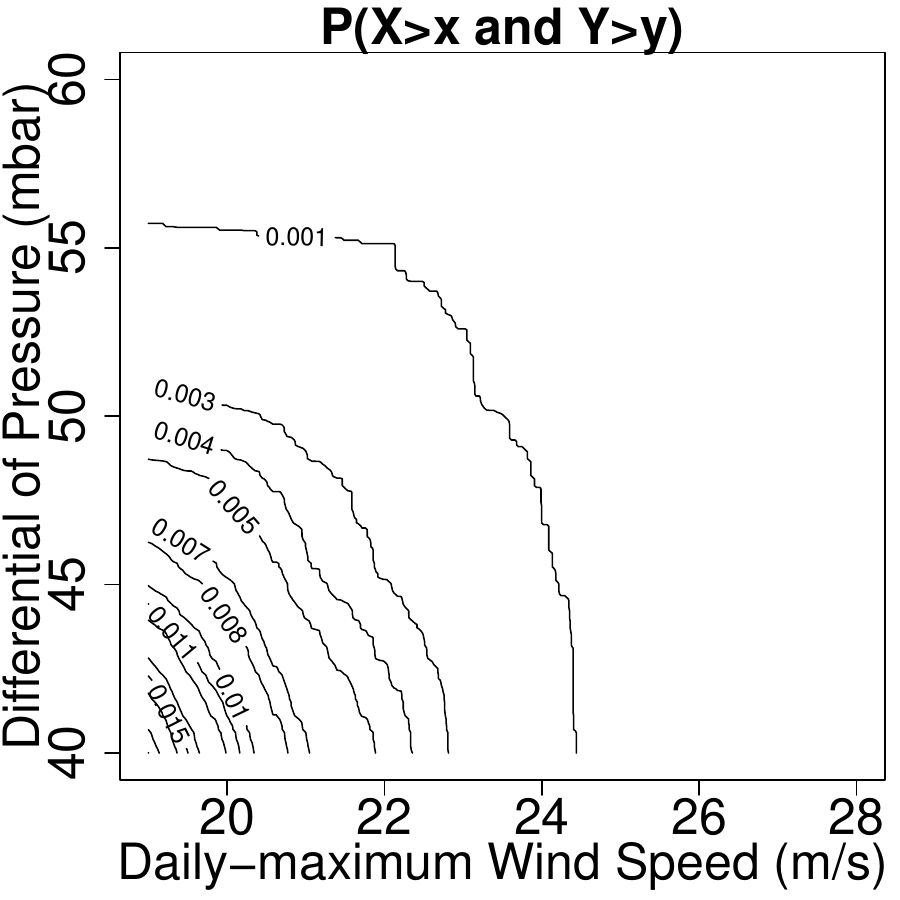}
\caption{\label{fig:pfailure} Estimated probabilities to belong to the failure regions $\mathcal{A}_{\boldsymbol{u}}$ (left) and $\mathcal{B}_{\boldsymbol{u}}$ (right), for $\boldsymbol{u}=(u_1, u_2)$, with $18 \leq u_1 \leq 28, 40 \leq u_2 \leq 60$.}
\end{figure}

The routine \code{pFailure} computes empirical estimates of probabilities of belonging to the failure sets $\mathcal{A}_{\boldsymbol{u}}$ (\code{type = "or"}) and $\mathcal{B}_{\boldsymbol{u}}$ (\code{type = "and}) by applying formula \eqref {eq:empirical_prob} on data generated by \code{rExtDep}. Both probabilities are computed when \code{type = "both"}. The argument \code{plot} offers the possibility to display contour plots when sequences of thresholds are provided through the arguments \code{u1} and \code{u2}. The code provided below
generates $N=50,000$ samples to estimate the probabilities of belonging to each failure sets for thresholds ranging from 19 to 28 for the wind speed and from 40 to 60 for the differential of pressure with outputs given in Figure \ref{fig:pfailure}.
\begin{CodeChunk}
\begin{CodeInput}
R> # Estimate probabilities to belong to failure regions "or"/"and"
R> pF <- pFailure(n = 50000, beta = SP_mle$Ahat$beta, 
+                 u1 = seq(from = 19, to = 28, length = 200),  
+                 u2 = seq(from = 40, to = 60, length = 200), 
+                 mar1 = pars.WS, mar2 = pars.DP, type = "both", 
+                 plot = TRUE, xlab = "Daily-maximum Wind Speed (m/s)", 
+                 ylab = "Differential of Pressure (mbar)", nlevels = 15)
\end{CodeInput}
\end{CodeChunk}
%


%
\subsubsection[Extreme quantile regions]{Estimation of extreme quantile regions}
\label{sssec:exQ}
\noindent\textbf{Theory}. As highlighted in Section \ref{sec:back_ground}, an important problem is to define a quantile region $\mathcal{Q}$ as in \eqref{eq:quantile_region}, given a very small probability $p$ to fall in it. Given a sample of size $N$, an extreme quantile region can be defined by the level set $\mathcal{Q}_N$ such that $\prob(\mathcal{Q}_N)=p$, where $p=p_N$ satisfies $p\to 0$ and $Np\to c>0$ as $N\to\infty$. For practical purposes, $\mathcal{Q}_N$ can be approximated by the region $\widetilde{\mathcal{Q}}_N$ given in \eqref{eq:app_extreme_region} which requires estimating the marginal parameters $\bftheta_j=(\mu_j,\sigma_j,\gamma_j)^\top$, $j=1,2$, the basic set $S$ and the density of the angular measure, allowing, in turn, to obtain an estimate of $\xi(S)$. \cite{beranger2021} discussed a Bayesian approach for the inference of $\widetilde{\mathcal{Q}}_N$ based on a polynomial angular measure in Bernstein form and the censored-likelihood approach \citep[e.g.,][Ch. 8]{beirlant2004}. Since such a methodology is similar to the one described in Section \ref{sssec:beed_np}, we refer to \cite{beranger2021} for a full description.

The methodology is illustrated on the dataset \code{Milan.winter}, which consists of air pollution levels recorded in Milan, Italy, over the winter period October 31st–February 28/29th, between December 31st 2001 and December 30th 2017. We focus here on the $\textrm{NO}_2$ and $\textrm{SO}_2$ pollutants and use the daily maximum temperature (\code{MaxTemp}) as covariate. 
The data is prepared before estimating extreme quantile regions using \code{fExtDep.np}.
\begin{CodeChunk}
\begin{CodeInput}
R> data(MilanPollution, package = "ExtremalDep")
R> data <- Milan.winter[, c("NO2","SO2", "MaxTemp")]
R> data <- data[complete.cases(data), ]
\end{CodeInput}
\end{CodeChunk}
A quadratic relationship between pollutants and maximum temperature is considered, and the $j$th marginal mean is specified as $\mu_j = \beta_{0,j} + \beta_{1,j} z + \beta_{2,j} t^2$, with $j=1,2$, where $t$ is the temperature level. 
The covariate matrix is defined as
\begin{CodeChunk}
\begin{CodeInput}
R> covar <- cbind(rep(1, nrow(data)), data[, 3], data[, 3]^2 )
\end{CodeInput}
\end{CodeChunk}
which will be provided to \code{fExtDep.np} through the arguments \code{cov1} and \code{cov2}. 
Since a polynomial angular measure in Bernstein form is considered, we specify a prior distribution for the polynomial degree $\kappa$ as a negative binomial (default) for $\kappa - 3$ with mean $6$ and variance $8$ and priors for the point masses $p_0$ and $p_1$ respectively as uniform distributions (default) on $[0, 0.2]$ and $[a,b]$, where $a$ and $b$ are defined as in the previous two sections.
These are the default prior distributions in the \code{fExtDep.np} routine, therefore the hyper-parameters are specified by 
\begin{CodeChunk}
\begin{CodeInput}
R> hyperparam <- list(mu.nbinom = 6, var.nbinom = 8, a.unif = 0, b.unif = 0.2)
\end{CodeInput}
\end{CodeChunk}
The prior for $p_0$ and $p_1$ are set in such a way to represent the belief that those point masses are likely to be small for these data. The mean and variance of the negative binomial prior for the polynomial degree are set to allow for higher degree polynomial modelling of the angular density if required.

As mentioned above, the estimation is performed through the \code{fExtDep.np} function where starting values for the marginal parameters $\boldsymbol{\theta}_j^{(0)} = \left( \beta_{0,j}^{(0)}, \beta_{1,j}^{(0)}, \beta_{2,j}^{(0)}, \sigma_{j}^{(0)}, \gamma_j^{(0)} \right)^\top$, $j=1,2$, the scaling parameter of the sampler on each margin and the polynomial degree are respectively set by the arguments  \code{par10}, \code{par20}, \code{sig10} \code{sig20} and \code{k0}. The argument  \code{method = "Bayesian"} indicates that a Bernstein polynomial is used to represent the extremal dependence and that the inference part follows the MCMC scheme detailed in \citet[][Algorithm 1]{beranger2021}. The argument \code{u = TRUE} specifies that a censored likelihood approach is applied on raw data with threshold \code{u} set to the marginal $90\%$ quantile by default. Recall that by default the marginal distributions are fitted jointly with the dependence (\code{mar.fit = TRUE}) but \code{mar.prelim = FALSE} indicates that no initial marginal fit is required. 
The code chunk below provides a full specification of the \code{fExtDep.np} arguments.

\begin{CodeChunk}
\begin{CodeInput}
R> pollut <- fExtDep.np(x = data[, -3], method = "Bayesian", u = TRUE, 
+    cov1 = covar, cov2 = covar, mar.prelim = FALSE, 
+    par10 = c(100, 0, 0, 35, 1), par20 = c(20, 0, 0, 20, 1), sig10 = 0.1, 
+    sig20 = 0.1, k0 = 5, hyperparam = hyperparam, nsim = 15e+3)
\end{CodeInput}
%\begin{CodeOutput}
%U set to 90% quantile by default on both margins
%
%Estimation of the extremal dependence and margins 
%
%  |====================================================================| 100%
%\end{CodeOutput}
\end{CodeChunk}

Numerical and graphical summaries of the output of the estimation procedure are obtained from the \code{summary} and \code{plot} functions by specifying the arguments \code{object}, \code{x}, \code{summary.mcmc} and \code{type = "Qsets"}. Since covariates are used to model the marginal parameters, we need to provide the covariate levels at which the extreme quantile regions will be computed. In this example, extreme quantile regions are computed at the minimum, median and maximum of the observed daily maximum temperatures. The covariate matrices \code{QatCov1} and \code{QatCov2} should not include an intercept term and contain a maximum of three levels as given through
\begin{CodeChunk}
\begin{CodeInput}
R> pollut_sum <- summary_ExtDep(mcmc = pollut, burn = 5e+3)
R> Temp.seq <- c(min(data[, 3]), median(data[, 3]), max(data[, 3]) )
R> QatTemp <- cbind(Temp.seq, Temp.seq^2)
\end{CodeInput}
\end{CodeChunk}
Representations of quantile regions associated with small probabilities $p=1/600, 1/1200$ and $1/2400$ are displayed using the \code{plot} function where we request to include graphical summaries of the extremal dependence as well as displaying the data (\code{dep=TRUE} and \code{data}). 
%
\begin{figure}[t!]
\centering
\includegraphics[width=\textwidth]{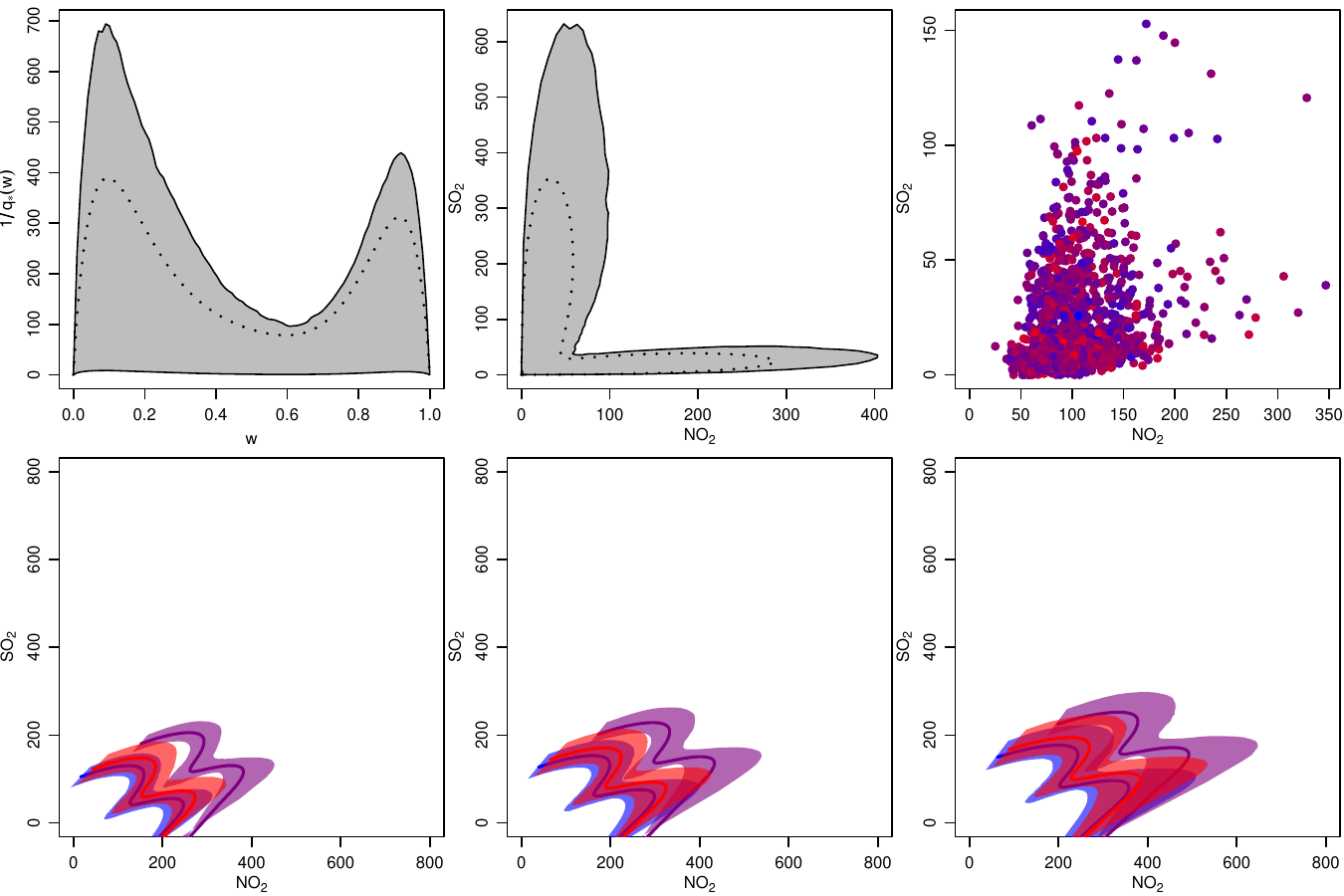}
\caption{\label{fig:pollut_qsets} Top row: posterior mean estimate (dotted line) and 90\% credible interval (grey) for the inverse of the angular basic density (left), the basic set $\mathcal{S}$ (middle), and observed data (right) with temperature dependent data colouring (from cold = blue to warm = red). Bottom row: posterior mean estimate (dotted line) and 90\% credible interval for the extreme bivariate quantiles associated with probabilities $p = 1/600, 1/1200$ and $1/2400$ (left to right) for three maximum daily temperature levels: minimum temperature (blue), median (purple) and maximum (red) temperatures.}
\end{figure}
The output 
contains a list of dependence quantities in \code{est.out} and a list of quantile sets in \code{q.out}. In particular, \code{est.out} includes the following elements: 
\begin{itemize}
\item \code{ghat}: a $3 \times 100$ matrix with an estimate of the inverse of the angular basic density $q_\star$ evaluated at 100 grid points in $[0,1]$. Rows gives the posterior $0.05$ quantile, mean and $0.95$ quantile. 
\item \code{Shat\_post} and \code{Shat}: lists where each element is a $100 \times 2$ matrix providing an estimate of the basic set $\mathcal{S}$ (obtained through \code{ghat}). \code{Shat\_post} considers every posterior samples whereas  \code{Shat} takes the pointwise $0.05$ quantile, mean and $0.95$ quantile.
\item \code{nuShat\_post} and \code{nuShat}: two vectors providing the posterior of the basic set measure $\xi(\mathcal{S})$ and its $0.05$ quantile, mean and $0.95$ quantile. 
\end{itemize}
The \code{q.out} list includes: 
\begin{itemize}
\item \code{Qset\_PA\_CovNum\_B}: a list of three $100 \times 2$ matrices. Each of them is an estimate of the bivariate extreme quantile $\widetilde{\mathcal{Q}}_n$. 
Such regions are computed for the specified probabilities (\code{P}) and covariate (\code{QatCov1} and \code{QatCov2}),
e.g., \code{pl\$Qset\_P1\_CovNum\_1} provides an estimate of the ($\text{NO}_2, \text{SO}_2$) region corresponding to the probability $1/600$, when the minimum temperature is observed. Each matrix corresponds to the posterior $0.05$ quantile, mean and $0.95$ quantile, obtained from \code{Qset\_PA\_CovNum\_B\_post}. 
\item \code{Qset\_PA\_CovNum\_B\_post}: a $2 \times$(\code{nsim-burn}) matrix providing an estimate of the extreme quantile region for each posterior sample.
\end{itemize}
%
An implementation is given below, where the argument \code{dep=TRUE} produces the top row of Figure~\ref{fig:pollut_qsets}. 
\begin{CodeChunk}
\begin{CodeInput}
R> pl <- plot(x = pollut, type = "Qsets", summary.mcmc = pollut_sum,  
+    QatCov1 = QatTemp, P = 1/c(600, 1200, 2400), dep = TRUE, 
+    data = data[, -3], xlim = c(0, 800),  ylim = c(0, 800), 
+    xlab = expression(NO[2]), ylab = expression(SO[2]))            
\end{CodeInput}
\end{CodeChunk}
The bottom row of Figure~\ref{fig:pollut_qsets} shows the extreme quantile regions corresponding to probabilities $1/600$ (left), $1/1,200$ (middle) and $1/2,400$ (right) where the colours indicates different levels of the covariate. Colours of the data (top right), credibility regions and mean from the posterior can be respectively specified by \code{col.data, col.Qfade} and \code{col.Qfull} (blue to red colour palette by default).

%
\section[Spatial extremes]{Spatial extremes}\label{sec_spa_extremes}
Max-stable processes \citep[e.g.,][Ch. 9]{de2006} are a convenient tool to statistically model extremes and their dependence over a spatial domain $\mathcal{D}$. Let $\{X_i(s), s\in \mathcal{D}\}$, $i=1,2,\ldots$, be independent copies of a stochastic process $X(s)$, $s\in\mathcal{D}$ with identical finite-dimensional distribution. If there are functions $a_n(s)>0$ and $b_n(s)$, for each $n$, such that the finite-dimensional distributions of a limit process $Y(s)$, given by
$$
\{Y(s),s\in \mathcal{D}\} = \lim_{n\to\infty}\left\{\max_{1\leq i\leq n} \left(\frac{X_i(s)-b_n(s)}{a_n(s)}\right), s\in\mathcal{D}\right\},
$$
are not degenerate, then they must be multivariate EV as in \eqref{eq:evd_copula}, and with the maximum taken pointwise for all $s\in \mathcal{D}$. Refer to \cite{davison2012statistical} for a detailed review of max-stable processes and the statistical analysis of spatial extremes. The \pkg{ExtremalDep} package considers some of the most widely used max-stable models including geometric Gaussian \citep{davison2012statistical}, Brown--Resnick \citep{Brown1977}, extremal-$t$ \citep{Opitz2013} and extremal skew-$t$ processes \citep{Beranger2017}. The routines implemented rely on the results of \citet{Dombry2016} and \citet{Beranger2021b} for the exact simulation of the max-stable models and on \citet{Stephenson2005} and \citet{Beranger2021b} for the inferential procedures.\\

\noindent\textbf{Software implementation}. The routine \code{rExtDepSpat} extends the \code{rmaxstab} function from the \pkg{SpatialExtremes} package \citep{SpatialExtremes210} by including exact and direct simulation (\code{method="direct"} or \code{"exact"}) from the extremal skew-$t$ process,  by using the prefix ``s'' when defining the type of correlation function in the argument \code{cov.mod} (options are \code{"whitmat"}, \code{"cauchy"}, \code{"powexp"} and \code{"bessel"}). For the extremal skew-$t$ model, the skewness parameter is represented as
$$
\boldsymbol{\alpha}  = \alpha_0 + \alpha_1 \textbf{cov}_1 + \alpha_2 \textbf{cov}_2 \in \mathbb{R}^d,
$$ 
where $\textbf{cov}_1$ and $\textbf{cov}_2$ are $d$-dimensional covariate vectors with $d$ the number of spatial locations $s_j, j=1,\ldots,d$. The argument \code{alpha} is used to specify $(\alpha_0, \alpha_1, \alpha_2)$ while the covariates are given by \code{acov1} and \code{acov2}. Using 
the command lines provided below,
$50$ (\code{Ny}) replicates of the extremal-$t$, with $\nu = 1$ degrees of freedom (\code{DoF}), are generated at $10$ (\code{Ns}) spatial locations (\code{sites}) in the region $[-5,5]^2$. The correlation function (\code{cov.mod}) is the power exponential class $\rho(h) = \exp \left\{ - \left( \| h \| / r \right)^\eta \right\}$ with smoothness (\code{smooth}) parameter $\eta=1.5$ and range (\code{range}) parameter $r=3$. 
\begin{CodeChunk}
\begin{CodeInput}
R> set.seed(14342)
R> Ns <- 10; Ny <- 50
R> sites <- matrix(runif(Ns*2) * 10 - 5, nrow = Ns, ncol = 2)
R> for(i in 1:2) sites[, i] <- sites[, i] - mean(sites[, i])

R> # Generate samples from the extremal-t model with power exp corr function
R> z <- rExtDepSpat(Ny, sites, cov.mod = "tpowexp", DoF = 1, range = 3, 
+    nugget = 0, smooth = 1.5, control = list(method = "exact"))
\end{CodeInput}
\end{CodeChunk}
The routine \code{rExtDepSpat} returns a list consisting of simulated values at specified locations (\code{\$vals}) and the hitting scenario (\code{\$hits}), both being \code{Ny} $\times$ \code{Ns} matrices. For a given row of \code{\$hits}, elements with the same value indicate block maxima that occurred at the same time (for illustrative purposes, one may think that the maxima were obtained from the same ``storm''). 

The \code{fExtDepSpat} procedure takes advantage of the availability of the time of occurrence of block maxima to fit the extremal-$t$ (\code{model = "ET"}) and skew-$t$ (\code{model = "EST"}) max-stable models using the Stephenson---Tawn likelihood \citep{Stephenson2005}. The correlation function is currently restricted to the power exponential, meaning that the parameters vectors to be estimated are respectively $\bftheta = (\nu, r, \eta)^\top$ and $\bftheta = (\nu, r, \eta, \alpha_0, \alpha_1, \alpha_2)^\top$. 
The corresponding parameters can be fixed through arguments \code{range}, \code{smooth}, \code{DoF} and \code{alpha} of \code{fExtDepSpat}, else are estimated. Note that, a vector of length $3$ must be provided for \code{alpha} and therefore a \code{NA} value leaves the corresponding parameter free, e.g., \code{alpha = c(0, NA, NA)} would fit the extremal skew-$t$ model with skewness $\boldsymbol{\alpha}  = \alpha_1 \textbf{cov}_1 + \alpha_2 \textbf{cov}_2 $. Initial values are provided in \code{par0} in vector form as \code{c(DoF, range, smooth, alpha0, alpha1, alpha2)}, where fixed or unnecessary parameters are simply omitted. Computations can also be sped-up by considering parallelization (\code{parallel = TRUE}) and specifying the number of cores (\code{ncores}) to be used. Arguments \code{args1} and \code{args2} are related to specifications of the Monte Carlo simulation scheme to compute the multivariate $t$ cumulative distribution function (cdf). These should take the form of lists including the minimum and maximum number of simulations used (\code{Nmin} and \code{Nmax}), the absolute error (\code{eps}) and whether the error should be controlled on the log-scale (\code{logeps}). \code{args1} refers to the $d-1$ dimensional cdfs terms required to compute the exponent function while \code{args2} focuses on the $d-m$ ($2 \leq m \leq d-1$) dimensional cdfs involved in the evaluation of its partial derivatives. When computing the log-likelihood function, the latter terms need to be evaluated on the log-scale, requiring fewer Monte Carlo simulations. In  
the following command lines,
the strategy is to set a higher number of simulations for the partial derivative terms since experiments have shown that they are more important than the terms in the exponent function \citep[see][]{Beranger2021b}. The \code{control}  argument offers additional control parameters for the optimization algorithm see \code{?optim} for more details.
\begin{CodeChunk}
\begin{CodeInput}
R> # Set min and max number of sim to compute CDF terms
R> args1 <- list(Nmax = 50L, Nmin = 5L, eps = 0.001, logeps = FALSE)
R> args2 <- list(Nmax = 500L, Nmin = 50L, eps = 0.001, logeps = TRUE)

R> # Fit true (extremal-t) model using full likelihood
R> fit1 <- fExtDepSpat(x = z$vals, model = "ET", sites = sites, hit = z$hits, 
+    par0 = c(3, 1, 1), parallel = TRUE, ncores = 2, args1 = args1, 
+    args2 = args2, control = list(trace = 0))
R> est(fit1)
\end{CodeInput}
\begin{CodeOutput}
  DoF range smooth 
0.991 2.940  1.551 
\end{CodeOutput}
\begin{CodeInput}
R> method(fit1)
\end{CodeInput}
\begin{CodeOutput}
[1] "Full likelihood"
\end{CodeOutput}
\end{CodeChunk}

The routine \code{fExtDepSpat} returns an object of class \code{ExtDep\_Spat} from which the estimated parameters, the estimation method, the standard errors computed from the sandwich information matrix, the value of the maximised log-likelihood  and the Takeuchi information criteria can respectively be extracted using the functions \code{est}, \code{method}, \code{StdErr}, \code{logLik} and \code{tic}.
Here the estimated parameter vector is $\widehat{\bftheta} = (0.991, 2.940, 1.551)^\top$, while the true parameters are $\bftheta=(1,3,1.5)^\top$ using the default full likelihood 
as indicated by \code{est1\$jw} taking value $20$.
As proposed in \citet{Beranger2021b}, composite likelihood estimation \citep{padoan2010likelihood} is incorporated when the argument \code{jw} is specified and is less than the number of locations. Since the number of tuples (pairs if \code{jw = 2}, triples if \code{jw = 3}) is increasing with \code{jw} one can specify a threshold $u$ through the argument \code{thresh} such that, for a tuple $q$, the corresponding composite likelihood contribution is weighted according to
$$
w_q = \left\{
\begin{array}{cl}
1 & \textrm{if } \max_{i,k \in q; i \neq k} \| s_i - s_k\| <u, \\
0 & \textrm{otherwise}.
\end{array}
\right.
$$
In other words, only tuples with maximum pairwise distance less than $u$ are included in the likelihood.
\begin{CodeChunk}
\begin{CodeInput}
R> # Fit true (extremal-t) model using composite likelihood
R> fit2 <- fExtDepSpat(x = z$vals, model = "ET", sites = sites, hit = z$hits, 
+    jw = 3, thresh = quantile(dist(sites), 0.25), par0 = c(3, 1, 1), 
+    parallel = TRUE, ncores = 2, args1 = args1, args2 = args2, 
+    control = list(trace = 0))
R> est(fit2)
\end{CodeInput}
\begin{CodeOutput}
  DoF  range smooth 
1.051  2.975  1.695  
\end{CodeOutput}
\begin{CodeInput}
R> method(fit2)
\end{CodeInput}
\begin{CodeOutput}
[1] "3-wise composite likelihood"
\end{CodeOutput}
\end{CodeChunk}
The above code script indicates that fitting the extremal-$t$ model using the triplewise composite-likelihood approach with threshold $u$ set at the first quartile of pairwise distances yields $\widehat{\bftheta} = (1.051, 2.975, 1.695)^\top$, which is again close to the true one.\\
%

\begin{figure}	
	\centering
	\includegraphics[width=0.4\textwidth]{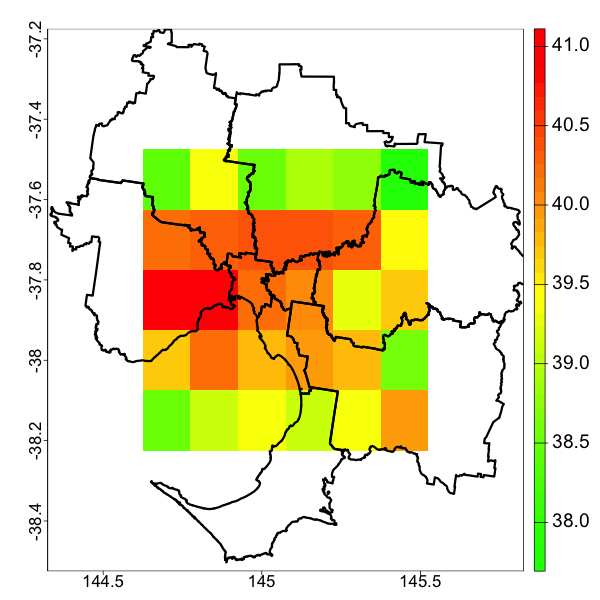}
	\caption{Simulation  (in $^\circ$C unit) from the fitted extremal skew-$t$ model with $\nu=5$, conditioning on at most two heatwave events causing all maxima.}
	\label{fig:condSim}
\end{figure}

\noindent\textbf{Application}. We now analyze temperature data around Melbourne, Australia, collected at $90$ stations on a $0.15$ degree (approximately $13$ kilometer) grid in a $9$ by $10$ formation, over the extended summer period from August to April between $1961$ and $2010$. Running the command \code{data(heat, package = "ExtremalDep")} loads several datasets in \pkg{R}. The objects \code{locgrid}, \code{scalegrid} and \code{shapegrid} are matrices of the marginal GEV parameters estimated over the grid using unconstrained location and scale while the shape parameter is defined as a linear function of eastings and northings in $100$ kilometer units. 
Below, the \pkg{terra} package \citep{Hijmans2024} is leveraged to provide graphical illustrations of these objects. 
\begin{figure}
	\begin{center}
		\includegraphics[width=5cm]{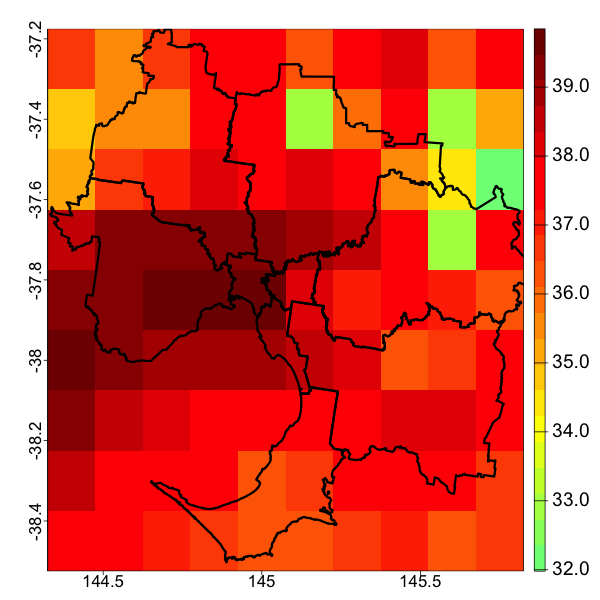}
		\hspace{-0cm}
		\includegraphics[width=5cm]{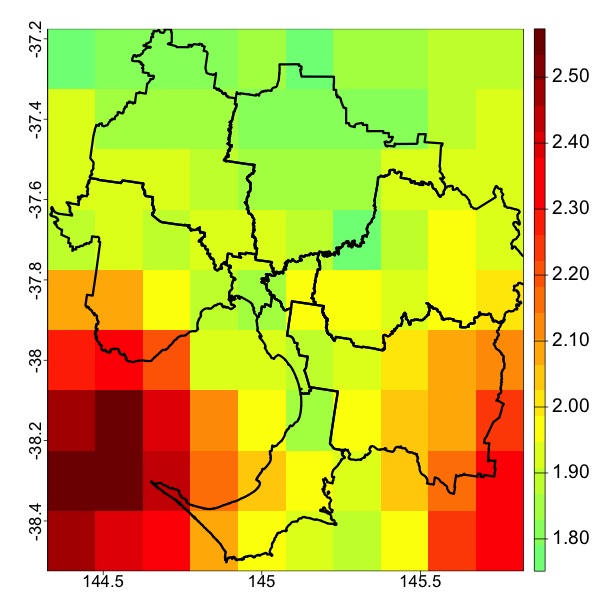}
		\hspace{-0cm}
		\includegraphics[width=5cm]{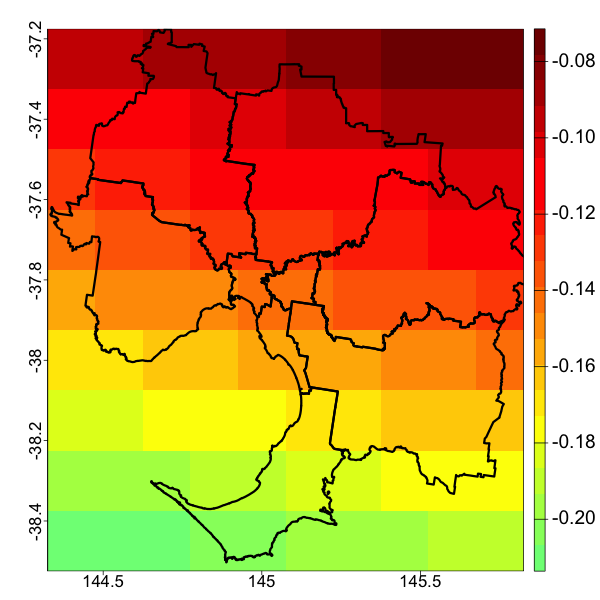}
	\end{center}
	\vspace{-0.5cm}
	\caption{Estimated marginal location (left), scale (middle) and shape (right) parameters.}
	\label{fig:gevPars}
\end{figure}
\begin{CodeChunk}
\begin{CodeInput}
R> library(terra)

R> data(heat, package = "ExtremalDep")

R> # Obtain boundary files for the study region
R> path <- "https://www.abs.gov.au/statistics/standards/australian-statistical-
+    geography-standard-asgs-edition-3/jul2021-jun2026/access-and-downloads/
+    digital-boundary-files"
R> file.name <- "SA4_2021_AUST_SHP_GDA2020.zip"
R> download.file(file.path(path, file.name), destfile = file.name)
R> unzip(file.name)
R> geogmel <- vect("SA4_2021_AUST_GDA2020.shp")
R> geogmel <- geogmel[geogmel$GCC_NAME21 == "Greater Melbourne"]
R> unlink(list.files(pattern = "SA4_2021_AUST"))

R> # Create grid and display marginal parameters
R> lat <- seq(from = -38.45, by = 0.15, length = 9)
R> lon <- seq(from = 144.4, by = 0.15, length = 10)
R> xx <- expand.grid(lon = lon, lat = lat)
R> cols <- tim.colors(40)[20:40]
R> for(m in c("locgrid", "scalegrid", "shapegrid")){
+    mat <- mget(m)
+    xx$data <- as.vector(mat[[1]])
+    rd <- rast(xx, crs = crs(geogmel))
+    plot(rd, col = map.pal("gyr", 40), col = cols, plg = list(cex = 2),  
+        pax = list(cex.axis = 2))
+    lines(geogmel, lwd = 3)
+  }
\end{CodeInput}
\end{CodeChunk}
Note that the code first downloads shape files and extracts the Greater Melbourne region.
Also provided in \code{heat} is the object \code{heatdata} which is a list with elements:
\begin{itemize}
\item \code{vals}: the $50$ yearly maxima (rows) at each of the $90$ locations (columns) 
\item \code{ufvals}: \code{vals} marginally transformed to unit-Fr\'{e}chet scale
\item \code{hits}: the hitting scenarios for each year (row) and location (row). Locations with the same integer value correspond to maxima obtained on the same day ($\pm3$ days). 
\item \code{sitesLLO}, \code{sitesENO}: original location coordinates in latitude-longitude (LL) and easting-northing (EN, in kilometer)
\item \code{sitesLL}, \code{sitesEN}: same as \code{sitesLLO}, \code{sitesENO} but centered.
\end{itemize}
Below, we fit the extremal skew-$t$ model with $\nu=5$ to the data. Since the observations were marginalized with GEV shape parameter as function of eastings and northings in 100 kilometre units, the site locations are transformed to be on the same scale. \cite{Beranger2021b} used the TIC for model selection between the extremal-$t$ and extremal skew-$t$ with $\nu=1,3$ and $5$ degrees but, for simplicity, we only consider here the selected model and a subset of the locations (\code{ind\_sites}).
\begin{CodeChunk}
\begin{CodeInput}
R> # Change to 100 kilometer units
R> sites <- heatdata$sitesEN/100

R> # Restrict to sub-region for faster computations
R> ind_sites <- (abs(sites[, 1]) < 0.4 & abs(sites[, 2]) < 0.4)
R> sites <- sites[ind_sites, ]

R> x <- heatdata$ufvals[, ind_sites]
R> hits <- heatdata$hits[, ind_sites]

R> args1 <- list(Nmax = 20L, Nmin = 2L, eps = 0.001, logeps = FALSE)
R> args2 <- list(Nmax = 200L, Nmin = 20L, eps = 0.001, logeps = TRUE)

R> # Fit extremal skew-t model
R> est5 <- fExtDepSpat(x = x, model = "EST", sites = sites, hit = hits, 
+    par0 = c(1, 1, 0.5, 0, 0), DoF = 5, acov1 = sites[, 1], 
+    acov2 = sites[, 2], parallel = TRUE, ncores = 2, args1 = args1, 
+    args2 = args2, control = list(trace = 2))
R> est(est5)               
\end{CodeInput}
\begin{CodeOutput}
  range  smooth alpha.0 alpha.1 alpha.2 
  9.275   1.262   0.496   0.001  -0.009 
\end{CodeOutput}
\end{CodeChunk}
The estimated range and smoothness of the skew-$t$ are respectively $\widehat{r}=9.275$, $\widehat{\eta}=1.262$, while the skewness is estimated as $\widehat{\boldsymbol{\alpha}} = 0.496 + 0.001 \textbf{easting} -0.009 \textbf{northing}$. The largest distance between pairs of locations (in 100 kilometer units) is $1.785$, and therefore the smallest correlation is $\exp\left[-(1.785/\hat{r})^{\hat{\eta}}\right] \approx 0.88$, indicating a strong degree of spatial dependence. Finally, the \code{rExtDepSpat} function is used to simulate from the extremal skew-$t$ conditionally on the hitting scenario and Figure~\ref{fig:condSim} displays an example conditioning on at most two heatwave events causing the annual maxima. 
These steps are implemented in the following code script.
\begin{CodeChunk}
\begin{CodeInput}
R> set.seed(123)
R> while(TRUE){
+     z.new <- rExtDepSpat(1, sites, cov.mod = "spowexp", DoF = 5, nugget = 0, 
+         range = est(est5)[1],  smooth = est(est5)[2], 
+         alpha = est(est5)[3:5], acov1 = sites[,1], acov2 = sites[,2], 
+         control = list(method = "exact"))
+     if(length(unique(as.numeric(z.new$hits))) <= 2) break
+  }

R> # Transform back to original margins
R> parmat <- cbind(as.vector(locgrid), as.vector(scalegrid), 
+    as.vector(shapegrid))
R> z.new.GEV <- rep(NA, 90)
R> z.new.GEV[ind_sites] <- trans2GEV(data = z.new$vals, 
+    pars = parmat[ind_sites, ])

R> xx$data <- z.new.GEV
R> rd <- rast(xx, crs = crs(geogmel))  
R> plot(rd, col = map.pal("gyr", 40), col = cols,  plg = list(cex = 2), 
+    pax = list(cex.axis = 2))
R> lines(geogmel, lwd = 3)          
\end{CodeInput}
\end{CodeChunk}

%
\section[Conclusions]{Conclusions}\label{sec_conclusion}

This paper has presented a comprehensive overview of methodological and practical modeling of multivariate and spatial problems.  Building on the componentwise maxima framework, we have shown how extremal dependence, characterized through angular measures and Pickands dependence functions, can be effectively estimated, interpreted, and exploited for risk assessment. A key message throughout the paper is that modeling extremal dependence is not only relevant for block maxima, but also provides a powerful basis for approximating probabilities of multiple threshold exceedances, joint tail events, and extreme quantile regions, thereby supporting a broad range of applied risk analyses.

The \pkg{ExtremalDep} package provides a flexible and extensible software that integrates parametric, semiparametric, and nonparametric approaches to modeling extremal dependence. Compared with existing software, it places particular emphasis on nonparametric and semiparametric representations of extremal dependence, to remain close to its nature, while still accommodating parametric models when interpretability or parsimony is required. In addition, it offers a Bayesian inference toolkit that allows for a principled and accessable uncertainty quantification. Through a series of real-world applications, ranging from air pollution and precipitation to financial returns, wind extremes, and heatwaves, the paper has demonstrated how the \pkg{R} package can be used to estimate extremal dependence, compute joint and conditional tail probabilities, derive multivariate return levels, simulate extreme events, and construct extreme quantile regions in both multivariate and spatial settings.

We plan the following future developments. An important extension concerns the integration of a more a general peaks-over-threshold inferential framework for multivariate data to better complement the  componentwise maxima. Scaling nonparametric and Bayesian approaches to even higher dimensions, for example through sparsity assumptions, graphical representations of extremal dependence, or dimension-reduction techniques. In the spatial domain, incorporating spatio-temporal dynamics and more flexible covariate effects into extremal dependence models remains a challenging avenue for research.
Future versions of \pkg{ExtremalDep} will aim to improve computational efficiency, for instance by exploiting parallel computing and more efficient Monte Carlo schemes, and to expand the range of implemented models and diagnostic tools and providing additional visualization and decision-oriented outputs.

\section*{Funding}\label{sec:acknowledgements}

BB is supported by the Australian Research Council through the Discovery Project Scheme (DP220103269).
SAP is supported by the Bocconi Institute for Data Science and Analytics (BIDSA) and MUR–PRIN
Bando 2022–prot. 20229PFAX5, financed by the European Union - Next Generation
EU, Mission 4 Component 2 CUP J53D23004260001.

\bibliographystyle{chicago}
\bibliography{ExtremalDep}

\appendix

\section{Appendix}

%
%
%
%
%
%

\subsection{Modelling of log-returns of exchange rates between GBP/USD and GPB/JPY: Goodness of fit }
\label{app:GoF}

\begin{figure}[h!]
\centering
\includegraphics[width=\textwidth]{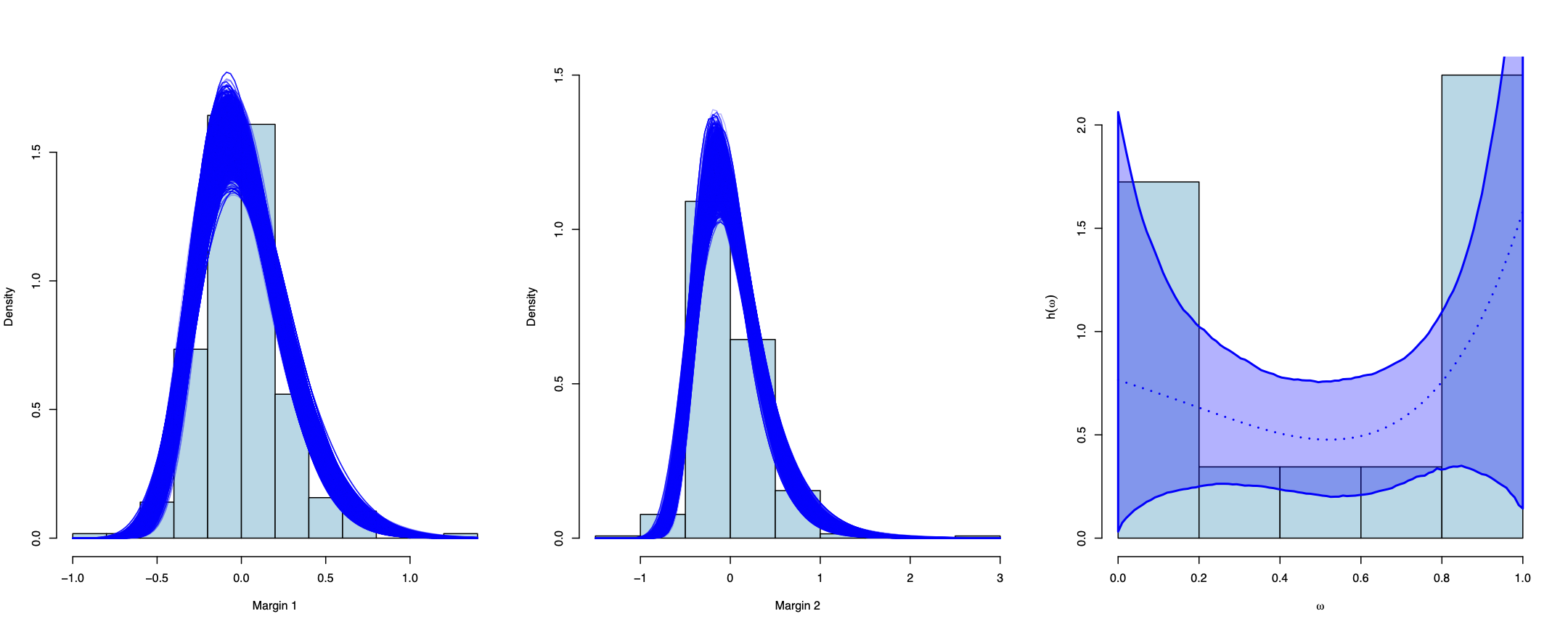} 
\caption{\label{fig:GBP_GoF} Left and middle panels: marginal density histograms (light blue) for GBP/USD and GPB/JPY with GEV density curves (solid blue) evaluated at each posterior sample for its corresponding parameter. Right panel: histogram of the empirical angular density $h(\omega)$ (light blue) with posterior mean of the estimated angular density (dotted blue) and $90\%$ credibility envelope.} 
\end{figure}

\subsection{Adaptive Gaussian random-walk Metropolis--Hastings (RWMH) algorithm}
\label{app:RWMHalg}

At the current state $s$ of the chain, $\bftheta_j^{(s)}$ is updated by the proposal 
$\bftheta_j'\sim \phi_3(\bftheta^{(s)}, \tau^{(s)} \Sigma^{(s)})$, where $\phi_d(\bfa, A)$ is a $d$-dimensional Gaussian density with mean $\bfa$ and covariance $A$. 
The proposal covariance matrix $\Sigma^{(s)}$ is specified as
\begin{equation}
\label{eq:A_update}
\Sigma^{(s+1)} = \left\{
\begin{array}{ll}
\left(1 + [\tau^{(s)}]^2 / s \right) \mathds{I}_3 , & s \leq 100 \\
\frac{1}{s-1} \sum_{l=1}^s \left(\bftheta^{(l)} - \bar{\bftheta}^{(s)}\right)\left(\bftheta^{(l)} - \bar{\bftheta}^{(s)}\right)^\top + \left\{ \left(\tau^{(s)}\right)^2/s\right\} \mathds{I}_3, & s > 100,
\end{array}
\right.
\end{equation}
where $\mathds{I}_d$ is the $d$-dimensional identity matrix, $\bar{\bftheta}^{(s)}=s^{-1}\left(\bftheta^{(1)}+\cdots+\bftheta^{(s)}\right)$,
and $\tau^{(s)} >0$ is a scaling parameter that affects the  acceptance rate of proposal parameter values \citep{haario+st01} and,
$\tau$ is updated using a Robbins--Monro process so that
\begin{equation}
	\label{eq:updatetau}
	\log \tau^{(s+1)} = \log \tau^{(s)} + c \left(\pi^{(s)} - \pi^*\right),
\end{equation}
where $c = (2\pi)^{1/2} \exp \left(\zeta_0^2/2 \right) / (2\zeta_0)$ is  a steplength constant, $\zeta_0 = - 1/\Phi(\pi^*/2)$, and $\Phi$ is the univariate standard Gaussian distribution function \citep{garthwaite2016}. To control the desired sampler acceptance probability, the parameter $\pi^*= 0.234$ is specified as in \citet{roberts1997}. We note that some improvements of the adaptive MCMC could be implemented to gain in efficiency, for example by taking as initial covariance a multiple of the inverse Fisher information. Furthermore, alternative strategies to guarantee diminishing adaptation could be implemented, e.g. by specifying a long burn-in period during which to adapt, then stop and start sampling.

\subsection{MCMC Scheme for the joint inference of marginal distribution and extremal dependence}
\label{app:MCMCalg}

\begin{algorithm}[b!]
\caption{Trans-dimensional MCMC scheme}
\label{alg:algo_joint}
\textbf{Initialize:} Set $\bfvartheta^{(0)} = \left(\bftheta_1^{(0)}, \bftheta_2^{(0)}, \kappa^{(0)}, \bfeta_{\kappa^{(0)}}^{(0)}\right),\bfeta_{\kappa^{(0)}}^{(0)}$, $\tau_i^{(0)}$ and $\Sigma_i^{(0)}$ for $j=1,2$. \;
\SetKwRepeat{REPEAT}{repeat}{until}
\For{$s = 0$ \To{} $M$}{
%
\textbf{Step 1:} {\em Marginal component 1:} \;
\quad Draw proposal $\bftheta_1' \sim \mathrm{MVN}(\bftheta_1^{(s)}, \tau^{(s)}_1 \Sigma^{(s)}_1)$.\;
\quad Compute acceptance probability 
$\pi_1 = \min \left\{  
 \frac{\Lik \left(\bftheta_1', \,\bftheta_{2}^{(s)}, \,\kappa^{(s)}, \,\bfbeta_{\kappa^{(s)}}^{(s)} \right)\Pi(\bftheta_1') }
       	{\Lik \left(\bftheta_1^{(s)},\, \bftheta_{2}^{(s)},\, \kappa^{(s)},\, \bfbeta_{\kappa^{(s)}}^{(s)} \right)\Pi\left(\bftheta_1^{(s)}\right) }, 1 \right\}.$\;
       	\quad Draw $U_1 \sim \mathcal{U}(0,1)$. If $\pi_1 < U_1$ then set $\bftheta_1^{(s+1)} = \bftheta_1'$ else set $\bftheta_1^{(s+1)} = \bftheta_1^{(s)}$.\;
\quad Update $\Sigma^{(s)}_1$ according to \eqref{eq:A_update}. \;
\quad Update $\tau^{(s)}_1$ according to \eqref{eq:updatetau}. \;
\textbf{Step 2:} {\em Marginal component 2:} \;
\quad Draw proposal $\bftheta_2' \sim \mathrm{MVN}(\bftheta_2^{(s)}, \tau^{(s)}_2\Sigma^{(s)}_2)$.\;
       	\quad Compute acceptance probability 
       	$\pi_2 = \min \left\{  
       	\frac{\Lik \left(\bftheta_1^{(s+1)},\, \bftheta_{2}',\, \kappa^{(s)},\, \bfbeta_{\kappa^{(s)}}^{(s)} \right)\Pi(\bftheta_2') }
       	{\Lik \left(\bftheta_1^{(s+1)},\, \bftheta_{2}^{(s)},\, \kappa^{(s)},\, \bfbeta_{\kappa^{(s)}}^{(s)} \right)\Pi\left(\bftheta_2^{(s)}\right)} , 1 \right\}.$\;
       	\quad Draw $U_2 \sim \mathcal{U}(0,1)$. If $\pi_2 < U_2$ then set $\bftheta_2^{(s+1)} = \bftheta_2'$ else set $\bftheta_2^{(s+1)} = \bftheta_2^{(s)}$.\;
        \quad Update $\Sigma^{(s)}_2$ according to \eqref{eq:A_update}. \;
        \quad Update $\tau^{(s)}_2$ according to \eqref{eq:updatetau}. \;
       \textbf{Step 3:} {\em Dependence structure:} \;
       \quad Draw proposal $\kappa' \sim q_\kappa(\kappa \mid \kappa^{(s)})$ and $\bfeta_{\kappa'}' \sim q_\eta(\bfeta_\kappa \mid \kappa')$, and compute $\bfeta_{\kappa'}'$.\;
\quad Set $c = 1/2$ if $\kappa^{(s)}=3$ or $c = 1$ if $\kappa^{(s)}>3$.\;
\quad Compute acceptance probability 
$\pi_3 = \min \left\{  c
       \frac{\Pi (\kappa')}{\Pi (\kappa^{(s)})}
       \frac{\Lik \left(\bftheta_1^{(s+1)},\, \bftheta_2^{(s+1)},\, \kappa',\, \bfbeta_{\kappa'}' \right) }
       {\Lik \left(\bftheta_1^{(s+1)},\, \bftheta_2^{(s+1)},\, \kappa^{(s)},\, \bfbeta_{\kappa^{(s)}}^{(s)} \right) } , 1 \right\}.$\;
       \quad Draw $U_3 \sim \mathcal{U}(0,1)$. If $\pi_3 < U_3$ then set $\kappa^{(s+1)}=\kappa', \bfeta_{\kappa^{(s+1)}}^{(s+1)}=\bfeta_{\kappa'}'$, $\bfbeta_{\kappa^{(s+1)}}^{(s+1)}=\bfbeta_{\kappa'}'$ else set $\kappa^{(s+1)}=\kappa^{(s)}, \bfeta_{\kappa^{(s+1)}}^{(s+1)}=\bfeta_{\kappa^{(s)}}^{(s)}$, $\bfbeta_{\kappa^{(s+1)}}^{(s+1)}=\bfbeta_{\kappa^{(s)}}^{(s)}$. \;
}
\end{algorithm}

\end{document}